\documentclass[manuscript,onecolappendix]{aastex631}

\usepackage{amsmath}
\usepackage{natbib}

\graphicspath{{./}{figures/}}

\begin{document}

\title{Modified Newtonian Dynamics as an Alternative to the Planet Nine Hypothesis}

\author[0000-0002-1567-5620]{Katherine Brown}
\affiliation{Physics Department, Hamilton College \\
198 College Hill Road \\
Clinton, New York, 13323  USA}

\author[0000-0003-4451-8963]{Harsh Mathur}
\affiliation{Physics Department, Case Western Reserve University \\
10900 Euclid Avenue \\
Cleveland, Ohio 44106 USA}
\collaboration{2}{}
\begin{abstract}

A new class of Kuiper belt objects that lie beyond Neptune with semimajor axes greater than 250 astronomical units show orbital anomalies that have been interpreted as evidence for an undiscovered ninth planet. We show that a modified gravity theory known as MOND (Modified Newtonian Dynamics) provides an alternative explanation for the anomalies using the well-established secular approximation. We predict that the major axes of the orbits will be aligned with the direction towards the galactic center and that the orbits cluster in phase space, in agreement with observations of  Kuiper belt objects from the new class. Thus MOND, which can explain galactic rotation without invoking dark matter, might also be observable in the outer solar system.

\end{abstract}


\vspace{5mm}


\section{Introduction} \label{sec:intro}

An exciting development in outer solar system studies is the discovery of a new class of Kuiper belt objects with orbits that lie outside that of Neptune and have semimajor axes in excess of 250 astronomical units \citep{sedna, trujillo, review}. The alignment of the major axes of these objects and other orbital anomalies are the basis for the hypothesis that a planet about 5-10 times as massive as Earth orbits the sun at an average distance of 500 astronomical units (au) \citep{batygin2016, review, bb1, bb2, bb3, malhotra}. Here we argue that a modified gravity theory known as MOND (Modified Newtonian Dynamics) \citep{milgrom1983, stacyreview, banikreview}  provides an alternative explanation for the observed alignment, owing to significant quadrupolar and octupolar terms in the MOND galactic field within the solar system \citep{milgromefe} that are absent in Newtonian gravity. We show using the well-established secular approximation \citep{elliotplusmcdermott} that MOND predicts  a population of Kuiper belt objects with orbits whose major axes are aligned along the direction towards the center of the galaxy and with aphelia oriented towards the galactic center. Moreover this population is predicted to cluster in phase space: the orbits should have high eccentricity and a propensity for their minor axes to be perpendicular to the direction to the center of the galaxy. All of these features are exhibited by known Kuiper belt objects belonging to the newly discovered class, in support of the MOND hypothesis. 
MOND was originally developed to explain galaxy rotation and its predictions on the 
galactic scale have 
recently been subject to stringent observational tests \citep{rar,stacyefe}.
Progress has also been made in applying MOND on the cosmological scale \citep{skordis}.
Hence evidence of MOND on solar system scales would further strengthen the case for it,
and establish the Kuiper belt as a laboratory for studying important questions of 
fundamental physics.


In the quasilinear formulation of MOND \citep{quasimilgrom}, the Newtonian gravitational field ${\mathbf g}_N$ is modified with an interpolating function $\nu$ to produce a ``pristine'' field ${\mathbf g}_P = \nu ( g_N / a_0) {\mathbf g}_N$ where $a_0$ is a fundamental acceleration scale. 
The interpolating function may induce a curl on the pristine field; 
${\mathbf g}_Q$ is the curl-free part of ${\mathbf g}_P$ and it is the physical field: in the absence of other forces, a test mass would experience acceleration ${\mathbf g}_Q$.
The interpolating function obeys $\nu (x) \rightarrow 1$ as $x \rightarrow \infty$ so that Newtonian gravity is recovered when the gravitational field
is strong compared to $a_0$ and $\nu(x) \rightarrow x^{-1/2}$ as $x \rightarrow 0$
in the weak field regime to ensure consistency with galaxy rotation curves. 
We take $\nu = 1/[1 - \exp(-\sqrt{x})]$, the same form that is
used by \cite{lelli2017}; see also \cite{zhu2023}. This form has the right asymptotic behavior and is
consistent with astrophysical and solar system constraints. 
For a point mass $M$ the characteristic MOND radius is given by 
$R_M = (G M / a_0)^{1/2}$; this is the distance at which the field crosses
over from the strong field Newtonian behavior to a MOND regime. 
Using $a_0 = 1.2 \times 10^{-10}$ m/s$^2$ (the best fit to galaxy
rotation data \citep{rar}) and 
the mass of the sun $M = M_\odot$ we find the MOND radius for the sun
$R_M = 7000$ au.
This provides the first clue that MOND effects may be detectable in the Kuiper belt. 

\section{Galactic field and the phantom mass}

\begin{figure}
\includegraphics[width=3.0in]{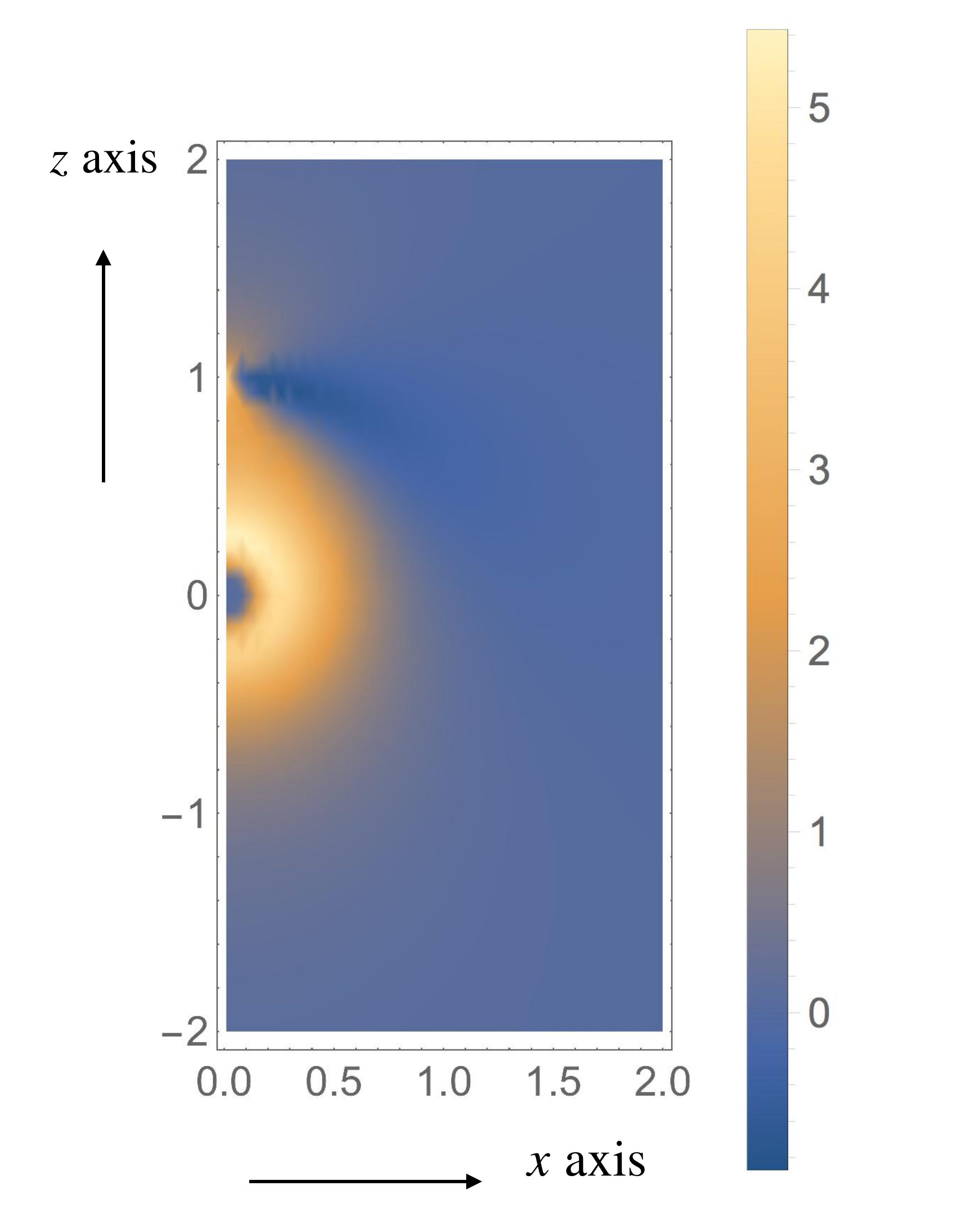}
\caption{{\em Phantom mass distribution.} 
The phantom mass is
rotationally symmetric about the $z$-axis and is localized at distances
comparable to the MOND radius $R_M$.
The sun is at the origin and
the galactic center is on the positive $z$-axis.  Distance along 
the axes is marked in units of $R_M = 7000$ A.U. and the
density $\rho_{{\rm ph}}$ is in units of $M_\odot/ (4 \pi R_M^3)$. 
We take $\gamma_N = 0.9 \; a_0$ consistent with the acceleration of
the sun towards the center of the galaxy \citep{stacymilkyway}. 
An interesting feature
is that the phantom mass distribution changes sign near the point 
where the Newtonian field of the galaxy and the sun add up to zero.
\label{fig:phantom}}
\end{figure}

In Newtonian gravity, the galactic field is essentially uniform on solar system scales and does not affect the observable relative motion of solar system bodies since they all accelerate equally in response to it. Tidal effects from nonuniformities in the galactic field are relevant to the Oort cloud but not the Kuiper belt \citep{oorttides}. The combined field of the sun at distance $r$ and the galaxy is thus
\begin{equation}
    {\mathbf g}_N = - \hat{{\mathbf r}} \frac{ G M_\odot}{r^2} + 
    {\boldsymbol \gamma}_N
    \label{eq:galacticnewton}
\end{equation}
where ${\boldsymbol \gamma}_N$ is the galactic field. 
In MOND however there is an additional anomalous field ${\mathbf g}_A$
within the solar system that can be interpreted as being due to a 
``phantom'' mass density $\rho_{{\rm ph}}$ \citep{milgromefe}
\begin{equation}
     \rho_{{\rm ph}} = - \frac{1}{4 \pi G}
    {\mathbf g}_N \cdot \nabla \nu ( g_N / a_0 );
    \label{eq:phantom}
\end{equation}
see Supplementary Information Section B (S.I. Sec B).
The relative motion of objects in the solar system is influenced by the anomalous MOND field ${\mathbf g}_A$, a phenomenon called the external
field effect \citep{milgromgalacticefe}
that has been recently observed on the galactic 
scale \citep{stacyefe, chae2021, chae2022, petersen2020, ascencio2022}. 
Unlike dark matter, the phantom mass has no independent dynamics; it is completely determined by the visible matter that produces ${\mathbf g}_N$. 
Fig. \ref{fig:phantom} shows the phantom mass distribution corresponding to the Newtonian field in Eq. (\ref{eq:galacticnewton}). Asymptotically, 
$\rho_{{\rm ph}}$
falls off as a power law for $r \gg R_M$ and
exponentially for $r \ll R_M$; there is essentially no phantom 
mass in the inner solar system.

We can compute the potential $\psi_A$ defined via ${\mathbf g}_A = 
- \nabla \psi_A$ using a multipole expansion (S.I. Sec C). 
To leading order the potential of a mass distribution localized far from the origin is given by the quadrupole term 
\begin{equation}
    \psi = - \frac{f G M}{R^3} r^2 P_2 (\cos \theta)
    \label{eq:quadrupole}
\end{equation}
where $f$ is a geometric factor, $M$ and $R$ the mass and length scales respectively that characterize the distribution, and 
$P_2$ is the second order Legendre polynomial. We have assumed the mass distribution is rotationally symmetric about the $z$-axis and $r \ll R$.
For the phantom mass distribution  $M = M_\odot$, $R = R_M$ and
$f = q_2 / 4 \pi$ where $q_2$ is a constant of order unity that depends
on $\gamma_N / a_0$ and the form of $\nu$. 
For our choice of $\nu$ and $\gamma_N / a_0 = 1.20$ we find $q_2 = 1.00$
by numerical integration. This value of $q_2$ is of the same order of 
magnitude as the Cassini bound \cite{hees2014} though it is numerically
larger than the bound. Other interpolating functions are shown by 
\cite{hees2016} to be consistent with the Cassini bound. Since here we 
are only interested in an order of magnitude comparison we may take
$q_2$ to be of order unity.

To compare the effect of this quadrupole field on a Kuiper belt object (KBO) with that of the hypothetical Planet Nine, we use the secular approximation \citep{elliotplusmcdermott}, wherein the mass of a planet is distributed nonuniformly along its orbit. The amount of mass on an arc of orbit is proportional to the time required for the planet to traverse the arc. 
The quadrupolar field of Planet Nine in this approximation is also given
by Eq. (\ref{eq:quadrupole}) with $M = m_9$ and $R = a_9$ (the mass and
semimajor axis respectively of Planet Nine) and $f = - (1 - e_9^2)^{-3/2}$
where $e_9 $ is the eccentricity of the Planet Nine orbit (S.I. Sec C). 
Note that the quadrupoles for MOND and Planet Nine are of opposite sign. 
Taking representative
values for the parameters from \cite{review} 
($m_9 = 5$ earth masses, $a_9 = 500$ A.U.
and $e_9 = 0.25$) we find that the orbit averaged quadrupole moment for
Planet Nine and MOND are the same order of magnitude. This provides 
further indication that MOND could have a significant effect on the orbits
of KBOs. 

\section{MOND and the outer Kuiper belt}

We now analyze the effects of the MOND field on the orbit of a KBO of mass $m_K$ using the secular approximation. Under the influence of the sun alone, a KBO would move along a Keplerian ellipse, with six orbital elements: semimajor axis $a_K$, eccentricity $e_K$, three Euler angles $(\omega_K, i_K, \Omega_K)$
and the ``mean anomaly'' which specifies where on the ellipse the KBO is located. 
We work in a frame with the sun at the origin and the center of the galaxy
located along the positive $z$-axis. In its reference configuration the KBO
orbit is assumed to lie in the $x$-$y$ plane with the perihelion on the
positive $x$-axis. The orientation $(\omega_K, i_K, \Omega_K)$ corresponds
to the orbit being rotated from the reference orientation successively about the $z$-axis by
$\omega_K$, the $x$-axis by $i_K$ and about the $z$-axis again by $\Omega_K$. 
In the secular approximation, the dynamics has two important simplifications \citep{elliotplusmcdermott}: $a_K$ is conserved, and the mean anomaly has no effect on the dynamics of the other orbital elements. The four remaining dynamical variables, $e_K$ and the Euler angles, undergo slow evolution
due to the MOND perturbation. 

In the secular approximation the dynamics of the orbit is controlled by a ``disturbing function'' which is the gravitational potential energy of the orbit-distributed mass of the KBO and the phantom mass distribution. This potential energy may be written in the form of a multipole expansion (S.I. Sec D). The disturbing function is often calculated in an approximate expansion in
eccentricity or inclination \citep{elliotplusmcdermott} but we need the quadrupole and octupole terms exactly. We have developed an efficient method to obtain these exact expressions.
To quadrupole order the secular disturbing function is 
\begin{equation}
    {\cal R}_Q = \frac{G m_K M_\odot}{R_M} 
    \left( \frac{a_K}{R_M} \right)^2
    \frac{q_2}{32 \pi} {\cal S}_Q
    \label{eq:disturbquad}
\end{equation}
where 
\begin{eqnarray}
    {\cal S}_Q & = & -2 - 3 e_K^2 + 15 e_K^2 \cos ( 2 \omega_K ) + 6 \cos^2 i_K
    \nonumber \\
& &    + 9 e_K^2 \cos^2 i_K - 15 e_K^2 \cos (2 \omega_K) \cos^2 i_K. 
    \label{eq:scaledquadrupole}
\end{eqnarray}
The octupole term and the derivation of the quadrupole and
octupole terms is given in S.I. Sec D. 

Due to the cylindrical symmetry of the anomalous field the secular disturbing function is independent of $\Omega_K$. As a result the component of the orbital angular momentum
along the symmetry axis, $(G M_\odot m_K a_K)^{1/2} h$ is conserved.
Here the scaled angular momentum 
\begin{equation}
    h = \sqrt{1 - e_K^2} \cos i_K.
    \label{eq:hdefinedone}
\end{equation}
Hence we have two conserved quantities, $h$ and ${\cal S}_Q$ itself, so 
the MOND dynamics of KBO orbits is integrable. We can eliminate $\cos i_K$
from ${\cal S}_Q$ using Eq. (\ref{eq:hdefinedone}) to obtain an expression for
${\cal S}_Q^{{\rm eff}}$ that depends only on $e_K$ and $\omega_K$. 

\begin{figure}
\includegraphics[width=6.in]{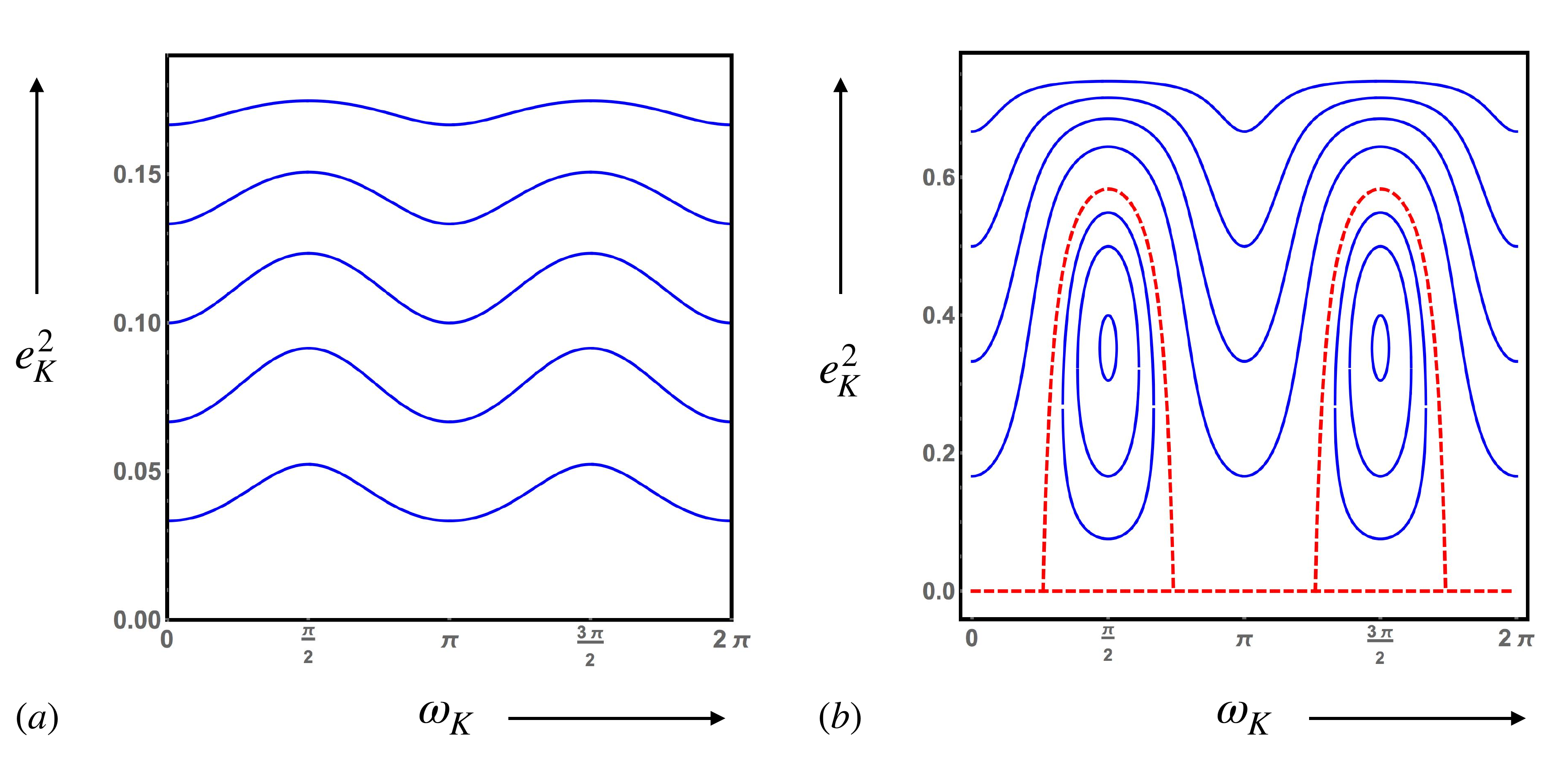}
 \caption{{\em Orbital dynamics in the quadrupole approximation.} 
 (a) Contours of fixed ${\cal S}^{{\rm eff}}_Q$ are wavy lines
 that extend across the $(e_K^2, \omega_K)$ plane for $3/5 \leq h^2 \leq 1$.
 (b) For $ 0 \leq h^2 < 3/5$ the phase space is dominated by two fixed points.
 The contours of fixed $S^{{\rm eff}}_Q$ are loops around the fixed points
 but remain wavy lines on the other side of the
 separatrix shown in red. The orbital elements thus precess along the
 wavy lines or oscillate about the fixed point values around the loops.
 The phase space flow is along the contours clockwise around the loops
 and from left to right (increasing $\omega_K$) along the wavy lines.
 The time scale of the dynamics is $10^7$ years for a KBO orbit 
 with a semimajor axis of 1000 au.
 \label{fig:contours}}
 \end{figure}

The orbital evolution can be visualized by plotting contours of fixed 
${\cal S}^{{\rm eff}}_Q$ 
in the $(e_K, \omega_K)$ phase plane. 
Fig. \ref{fig:contours} 
shows that for $0 \leq h^2 \leq 3/5$
the phase space is dominated by two fixed points located at 
$(e_C, \pi/2)$ and $(e_C, 3 \pi / 2)$ where $e_C^2 = 1 - \sqrt{5/3} | h |$. 
Over this range of $h$ the contours encircle the fixed points which are separated from a region of phase space where the contours are wavy lines extending across the phase plane. By use of Lagrange's equations we can show that the phase space flow is from left to right along the wavy lines (i.e. 
$\omega_K$ increases monotonically in time) while the loops are traversed clockwise (S.I. Sec E). The effect of the octupole term is to break the
symmetry between the fixed points: the one at 
$(e_C, \pi/2)$
becomes less stable while the one at 
$(e_C, 3 \pi / 2)$
becomes more stable. The dynamics of $i_K$ is dictated by Eq. (\ref{eq:hdefined}); 
and $\Omega_K$
undergoes precession (monotonic increase in time) 
regardless of the value of $|h|$ (S.I. Sec E).

Next we consider the effect of integrability-breaking terms on the orbital dynamics. These terms include non-secular terms in the MOND disturbing function and the secular and non-secular perturbations produced by the giant planets. Because of the long time scales involved it is also important to take into account the slow variation of the direction towards the center of the galaxy. According to Hamiltonian chaos theory \citep{percival} the phase space flow will become chaotic under the perturbations but a regular flow should persist around the stable fixed point near $(e_C, 3 \pi / 2)$,
particularly for small values of $h$. (For a given $a_K$ there is a minimum value of $|h|$ needed to ensure that the orbit does not penetrate the inner solar system; see S.I. Sec E). 

Thus we arrive at our central result: we predict the existence of a population of KBOs that are
localized near the fixed point $(e_C, 3 \pi / 2)$.
For small $h$ these orbits have inclinations $i_K \approx \pi / 2$. 
For an orbit with $\omega_K = 3 \pi / 2$ 
the apsidal vector $\hat{{\boldsymbol \alpha}}_K$
(the unit vector that points from the sun to the perihelion), then makes a small angle $\pi / 2 - i_K$
with $- \hat{{\mathbf n}}_G$, where $\hat{{\mathbf n}}_G$ 
is the unit vector that points from the sun to the center of the galaxy. Thus we
see that the apsidal vectors are aligned along the direction $-\hat{{\mathbf n}}_G$.
Finally, note that due to precession in $\Omega_K$ the orbits could have very high inclinations relative to the ecliptic. 

\section{Comparison to data}

\begin{figure}
 \includegraphics[width=3.5in]{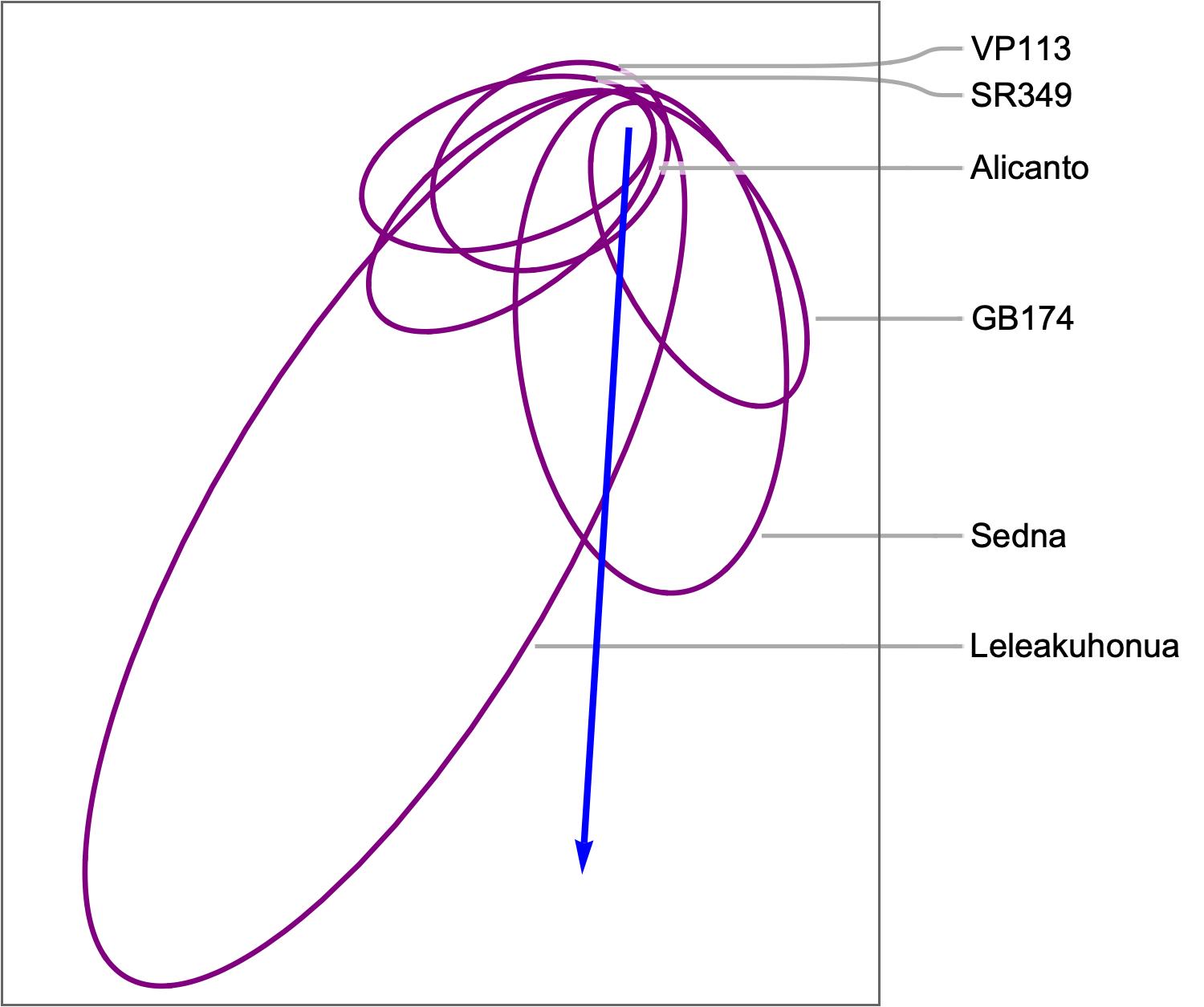}
 \caption{{\em Orbital alignment.} Orbits of six KBOs of the Sedna family 
 projected onto the ecliptic plane. The blue line is drawn parallel to the
 projection of $\hat{{\mathbf n}}_G$ onto the ecliptic plane. 
 The orbits and $\hat{{\mathbf n}}_G$ lie essentially in the ecliptic
 plane so the projection does not induce significant distortion. 
 The major axes of the orbits are seen to align with
 the direction to the center of the galaxy, with the aphelion 
 oriented towards the galactic center. This result is intuitively plausible
 since the aphelion is the heavy end of the orbit in secular
 perturbation theory. The KBO orbital parameters used to generate
 this plot are taken from the Minor Planet Database of the International
 Astronomical Union. 
 \label{fig:elliptical}}
 \end{figure}

We now examine to what extent KBOs of the newly discovered class (which we refer to as the Sedna family) conform to these predictions. \cite{review} identify six members of the Sedna family that have stable orbits under the influence of the known planets, and eight more that are metastable. The stable KBOs are a particularly good testing ground for the Planet Nine hypothesis as well as the MOND hypothesis advanced here.

Fig. \ref{fig:elliptical} 
shows the six Sedna family orbits that are stable, 
projected onto the ecliptic plane.
The well documented alignment of the orbits \citep{bb3}
is evident but the figure shows that the orbits are also
aligned with the direction to the center of the galaxy. 
This observation has not been noted before and is a prediction of the MOND hypothesis. To quantify the alignment we calculate 
$\hat{{\boldsymbol \alpha}}_K \cdot \hat{{\mathbf n}}_G$
for the six stable KBOs, and find a mean value of $-0.68$.
The expected value of this quantity is zero in the absence of MOND as there is otherwise no physical basis for a correlation between the apsidal vectors and the direction to the center of the galaxy. If we make the null hypothesis that the apsidal vectors of the KBOs are independent random variables uniformly distributed over the unit sphere then the observed value of $-0.68$ is 
$3 \sigma$
away from the expected value of zero. 
Alternatively, we note that according to the null hypothesis the quantity 
$u = (1 + \hat{{\boldsymbol \alpha}}_K \cdot \hat{{\mathbf n}}_G)/2$ is a random variable
uniformly distributed over the unit interval $0 \leq u \leq 1$. The six observed values of 
$u$ deviate strongly from a uniform distribution. The Kolmogorov-Smirnov test shows
that the observed deviation from a uniform distribution has a
cumulative probability of only 0.0045; hence the null hypothesis is falsified at a 
high level of confidence (see S.I. Section F). 
Our model does not require perfect antialignment between 
$\hat{{\boldsymbol \alpha}}_K$ and $\hat{{\mathbf n}}_G$;
a crude estimate suggests that $A_K$,
the angle between 
$\hat{{\boldsymbol \alpha}}_K$ and $\hat{{\mathbf n}}_G$,
should lie in the range $130^\circ$-$180^\circ$ compatible with
the data (S.I. Section F). 

Sometimes the longitude of perihelion $\varpi = \omega + \Omega$
is used as a proxy for apsidal alignment. However $\varpi$
correlates with alignment only for small inclinations and it is
a frame dependent quantity. We prefer to quantify the alignment
by calculating $\hat{{\boldsymbol \alpha}}_K \cdot \hat{{\bf n}}_G$ 
because it is an invariant quantity and is unambiguously a measure of alignment. 

\begin{figure}
 \includegraphics[width=4.0in]{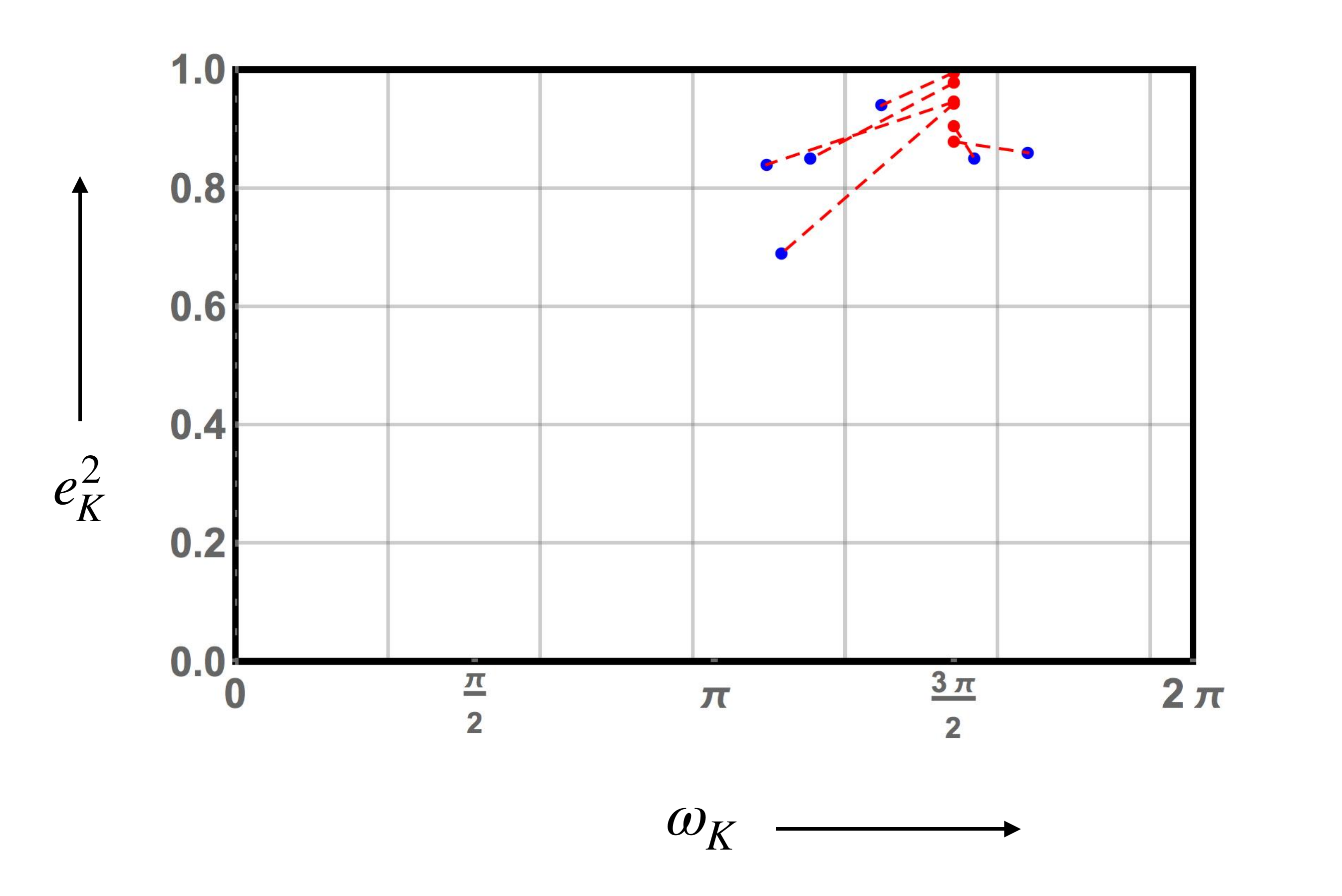}
 \caption{{\em Phase space clustering.} Plot of $(e_K^2, \omega_K)$ for the six KBOs of the
 Sedna family (blue points) and the location of the corresponding fixed
 point of the MOND orbital dyanmics in the quadrupole approximation
 (red points). The two points for each KBO have been connected by 
 a dashed line to guide the eye. 
 The KBO orbital parameters used to generate
 this plot are taken from the Minor Planet Database of the International
 Astronomical Union. 
 \label{fig:phasecluster}}
 \end{figure}

Fig. \ref{fig:phasecluster} shows the orbital elements $(e_K^2, \omega_K)$
as well as the location of the fixed point $(e_C^2, 3 \pi/ 2)$ 
for the six stable KBOs (See S.I. Section F for details).
As expected there is a clustering of the fixed points and KBO orbital elements.

Thus far we have concentrated on the six KBOs of the Sedna family that are known to have stable orbits under the influence of the known planets and that are therefore the best exemplars of the class. The review of \cite{review} identifies eight more KBOs that have metastable orbits and since the publication of the review eight additional KBOs that belong to the class have appeared in the Minor Planet Database. In S. I. Sec F we show that the alignment and phase space clustering shown in Figs 3 and 4 persists when all 22 KBOs are included in the plots. 

\section{Conclusion}

A larger sample of stable KBOs in the Sedna family compared to dynamical 
simulations would allow further tests of the MOND hypothesis.
The prospects for discovery of more Sedna-like objects are good. 
Existing surveys such as 
Dark Energy Survey \citep{des1, des2}, Transiting Exoplanet Survey Satellite \citep{tess} and Outer Solar System Origins Survey \citep{ossos1, ossos2}
as well as the forthcoming Vera Rubin Telescope \citep{schwamb1, schwamb2}
and CMB-S4 \citep{cowan, cmbs4}
all have the requisite sensitivity.
In addition to KBOs fast spacecraft are another promising probe of MOND in the outer solar
system \citep{banik2019, penner2020}. 

Historically, claimed gravitational anomalies in the solar system have almost invariably proven to be spurious under closer examination, albeit sometimes involving more than a century of debate and meticulous observation \citep{review, voyager,turyshev2012}. But they have also led to the discovery of Neptune and helped establish general relativity. It is possible that the Kuiper belt anomalies are evidence of Planet Nine, or that they are spurious \citep{lawler1, lawler2, bernardinelli2021, napier};
alternatively they may be evidence for a modification of Newtonian gravity.

\vspace{3mm}

We thank Stacy McGaugh for helpful discussions and the two anonymous reviewers
for suggesting references, helping improve the clarity of presentation and recommending 
the use of the Kolmogorov-Smirnov test. 

\bibliography{sample631}{}
\bibliographystyle{aasjournal}

\onecolumngrid

\newpage

\appendix

\section{Review of MOND field equations} 

The gravitational field is described by a vector ${\mathbf g}$ that corresponds
to the acceleration that would be experienced by a test mass due to gravity 
if no other forces were present. Newton's law of gravitation
may then be written as the pair of field equations 
\begin{equation}
    \nabla \cdot {\mathbf g}_N = - 4 \pi G \rho \hspace{3mm} {\rm and} 
    \hspace{3mm} \nabla \times {\mathbf g}_N = 0.
    \label{eq:newtonianfieldeq} 
\end{equation}
Here ${\bf g}_N$ is the Newtonian gravitational field produced by the mass
density $\rho$ and $G$ is Newton's gravitational constant. 

In the quasilinear formulation of MOND \citep{quasimilgrom}, 
given a mass distribution 
$\rho$, the first step is to calculate the Newtonian field ${\mathbf g}_N$
corresponding to that mass distribution. Next one calculates the pristine
field which is related to the Newtonian field via the nonlinear relation
\begin{equation}
    {\mathbf g}_P = {\mathbf g}_N \nu \left( \frac{g_N}{a_0} \right).
    \label{eq:pristine}
\end{equation}
Here $a_0 = 1.2 \times 10^{-12}$ m/s$^2$ is the MOND acceleration scale. The value
we have taken for this parameter is the recent best fit to galaxy rotation data \citep{rar}
but the value has in fact remained remarkably stable for decades \citep{begeman1991,gentile2011}.

The interpolating function $\nu$ has the asymptotic behavior 
\begin{equation}
    \nu (x) \rightarrow 1 \hspace{2mm} {\rm for} \hspace{2mm} x \gg 1 
    \hspace{3mm} {\rm and} \hspace{3mm}
    \nu (x) \rightarrow \frac{1}{\sqrt{x}} \hspace{2mm} {\rm for} \hspace{2mm} x \ll 1.
    \label{eq:asymptotic}
\end{equation}
The $x \gg 1$ behavior is fixed by the requirement that at strong fields MOND should
reduce to Newtonian gravity and the $x \ll 1$ behavior by the requirement that 
the field of a point mass should fall off as $1/r$ at large distances. 
For specific computations in this paper we have worked with the interpolating
function
\begin{equation}
    \nu (x) = \frac{1}{1 - \exp ( - \sqrt{x} )}.
    \label{eq:interpolfun}
\end{equation}
This form is consistent with galaxy rotation data and also with solar system
constraints arising from the ephemerides of the known planets. 

The next step is to compute the quasilinear field ${\mathbf g}_Q$ which is the
curl free part of ${\mathbf g}_P$. In MOND 
the quasilinear field ${\mathbf g}_Q$ is the
physical field: it is the acceleration 
experienced by test masses that are placed in the gravitational
field. 
It is now helpful to define the mass density $\rho_{{\rm eff}}$
\begin{equation}
    - 4 \pi G \rho_{{\rm eff}} = \nabla \cdot {\mathbf g}_P.
    \label{eq:rhoeffdef}
\end{equation}
It then follows that ${\mathbf g}_Q$ obeys the field equations
\begin{equation}
    \nabla \cdot {\mathbf g}_Q = - 4 \pi G \rho_{{\rm eff}} \hspace{3mm} {\rm and} \hspace{3mm} 
    \nabla \times {\mathbf g}_Q = 0.
    \label{eq:gqeq}
\end{equation}
In other words ${\mathbf g}_Q$ is the Newtonian field produced by the mass distribution 
$\rho_{{\rm eff}}$. We may regard $\rho_{{\rm eff}}$ as the equivalent distribution of
dark and visible matter that would be needed to mimic the effects predicted by MOND. However MOND
effects cannot always be mimicked by a suitable distribution of dark matter since
$\rho_{{\rm eff}}$ is not necessarily positive. 

As a simple illustration consider the field of a point mass $M$ located at the origin.
The Newtonian field is given by 
\begin{equation}
    {\mathbf g}_N = - \frac{G M}{r^2} \hat{{\mathbf r}}
    \label{eq:pointnewton}
\end{equation}
and hence the pristine field is given by
\begin{equation}
    {\mathbf g}_P = - \frac{G M}{r^2} \nu \left( \frac{R_M^2}{r^2} \right) \hat{{\mathbf r}}.
    \label{eq:pointpristine}
\end{equation}
Here the MOND radius $R_M$ is given by
\begin{equation}
    R_M = \sqrt{ \frac{GM}{a_0} }.
    \label{eq:mondradius}
\end{equation}
From the asymptotic forms of the interpolating function it follows that the pristine 
field has the asymptotic behavior
\begin{eqnarray}
    {\mathbf g}_P & \approx & - \frac{GM}{r^2} \hat{{\mathbf r}} \hspace{2mm} {\rm for} \hspace{2mm} r \ll R_M
    \nonumber \\
     & \approx & - \frac{GM}{R_M} \frac{1}{r} \hat{{\mathbf r}} \hspace{2mm} {\rm for} 
    \hspace{2mm} r \gg R_M.
    \label{eq:asymptoticpointmass}
\end{eqnarray}
Since ${\mathbf g}_P$ is radial and spherically symmetric it is also curl free. Hence in this
case the quasilinear field ${\mathbf g}_Q = {\mathbf g}_P$.
By computing the divergence of ${\mathbf g}_P$ we find that the equivalent dark matter density 
$\rho_{{\rm eff}}$ is a spherical halo that surrounds the point mass. The halo is peaked near
the MOND radius. For the interpolating function in eq (\ref{eq:interpolfun}) there is essentially
no mass density for $ r < R_M$ but there is a long tail as $r \rightarrow \infty$. 
For $M = M_{\odot}$ and the value of $a_0$ quoted above we find $R_M = 7000$ au, the MOND
radius of the sun. 

Finally, we note that we have chosen to use the quasilinear formulation
of MOND because of its simplicity. However essentially equivalent results are obtained
by use of the nonlinear formulation of MOND originally introduced by 
\cite{milgromgalacticefe}.

\section{Galactic field effect on the solar system}

Now let us consider the effect of the galactic gravitational field on the
solar system. On the scale of the solar system the galactic gravitational field
can be treated as essentially uniform. Ignoring all solar system objects except
the sun we must therefore solve
\begin{equation}
    \nabla \cdot {\mathbf g}_N = - 4 \pi G M_\odot \delta ({\mathbf r}) 
    \hspace{3mm} {\rm and} \hspace{3mm} \nabla \times {\mathbf g}_N = 0
    \label{eq:solarnewton}
\end{equation}
subject to the boundary condition ${\mathbf g}_N \rightarrow {\boldsymbol \gamma}_N$
as $r \rightarrow \infty$
where ${\boldsymbol \gamma}_N$ is the constant Newtonian field of the galaxy. 
The solution is of course
\begin{equation}
    {\mathbf g}_N = - \frac{G M_\odot}{r^2} \hat{{\mathbf r}} + {\boldsymbol \gamma}_N.
    \label{eq:newtonexternalfield}
\end{equation}
In Newtonian gravity, to the extent that the galactic field can be treated as uniform, 
it has no effect on 
the relative motion of bodies within the solar system since they all accelerate uniformly
in response to it. Tidal effects associated with variation of the galactic field are
important only for the most remote solar system objects in the Oort cloud. 
Remarkably we will now show that due to the nonlinearity of the equations, in MOND
the galactic field has an effect even on the inner solar system, a phenomenon 
dubbed the external field effect \citep{milgromefe, famaey2007, banik2018}. 

To analyze this problem within quasilinear MOND 
we make use of eq (\ref{eq:pristine}) and the product rule to obtain
\begin{eqnarray}
    \nabla \cdot {\mathbf g}_P & = & \left( \nabla \cdot {\mathbf g}_N \right) \nu \left( \frac{g_N}{a_0} \right)
    + {\mathbf g}_N \cdot \nabla \left[ \nu \left( \frac{g_N}{a_0} \right) \right]
    \label{eq:product}
\end{eqnarray}
where ${\mathbf g}_N$ is given by eq (\ref{eq:newtonexternalfield}). 
Evidently the first term on the right hand side is $- 4 \pi G M_\odot \delta ({\mathbf r})$.
Using eq (\ref{eq:rhoeffdef}), (\ref{eq:gqeq}) and (\ref{eq:product}) we obtain
\begin{equation}
    \nabla \cdot {\mathbf g}_Q = - 4 \pi G M_\odot \delta ({\mathbf r}) - 4 \pi G \rho_{{\rm ph}}
    \hspace{2mm} {\rm and} \hspace{2mm} \nabla \times {\mathbf g}_Q = 0.
    \label{eq:phantomeq}
\end{equation}
Here the phantom mass density $\rho_{{\rm ph}}$ is defined via 
\begin{equation}
    - 4 \pi G \rho_{{\rm ph}} = {\mathbf g}_N \cdot \nabla \left[ \nu \left( \frac{g_N}{a_0} \right)\right].
    \label{eq:phantomdef}
\end{equation}
Eq (\ref{eq:phantomeq}) must be solved subject to the boundary condition
${\mathbf g}_Q \rightarrow {\boldsymbol \gamma}_g$ for $r \rightarrow \infty$ where 
${\boldsymbol \gamma}_g$ is the physical acceleration due to gravity in the
solar system due to the galactic field. This is a measurable
parameter and we shall assume that ${\boldsymbol \gamma}_g = {\boldsymbol \gamma}_N \nu ( \gamma_N / a_0 )$.
The latter relation is a very good approximation for disc galaxies that do not have too flat an aspect
ratio \citep{brada, roshan}.

In order to analyze the effects of the galactic field on the relative motion of solar system 
bodies it is convenient to define the anomalous gravitational field ${\mathbf g}_A$ via
\begin{equation}
    {\mathbf g}_Q = {\mathbf g}_A - \frac{G M_\odot}{r^2} \hat{{\mathbf r}} + {\boldsymbol \gamma}_g.
    \label{eq:ganomalous} 
\end{equation}
${\mathbf g}_A$ captures effects beyond those due to the Newtonian field of the sun 
and the uniformly gravitational field of the galaxy. It embodies the external field effect. 

It is obvious from eqs (\ref{eq:phantomeq}) and  (\ref{eq:ganomalous}) that the anomalous field obeys
\begin{equation}
    \nabla \cdot {\mathbf g}_A = - 4 \pi G \rho_{{\rm ph}} \hspace{3mm} {\rm and} \hspace{3mm}
    \nabla \times {\mathbf g}_A = 0
    \label{eq:anomalousequation}
\end{equation}
subject to the boundary condition ${\mathbf g}_A \rightarrow 0$ and $r \rightarrow \infty$.
In short ${\mathbf g}_A$ is the Newtonian field produced by the phantom mass distribution. 

Eqs (\ref{eq:newtonexternalfield}), (\ref{eq:phantomdef}) and (\ref{eq:anomalousequation}) are the main results of this section. They provide an expression for the phantom mass that sources anomalous effects in the 
inner solar system. As discussed above these effects are absent in Newtonian gravity and hence
absent in any dark matter model. 


The external field effect within the solar system was analyzed by 
Milgrom within the original nonlinear MOND using 
a ``surrogate mass approximation''\citep{milgromefe}. Here we have 
followed that analysis closely but working within quasilinear MOND.
The main difference is that the analysis is simpler for quasilinear
MOND and it can be carried out exactly. This is not entirely by
chance: Milgrom constructed quasilinear MOND as a theory for which
the surrogate mass approximation would be exact \citep{quasimilgrom}. 

Finally for reference we provide a more explicit expression for the phantom mass.
It is convenient to work in a system of coordinates wherein the sun is at the origin 
and the galactic center along the positive $z$-axis. Then making use of eqs 
(\ref{eq:newtonexternalfield}) and (\ref{eq:phantomdef}) we obtain
\begin{equation}
    \rho_{{\rm ph}} = - \frac{M_\odot}{4 \pi R_M^3} \mu( \overline{r}, u )
    \label{eq:phantomexplicitone}
\end{equation}
where we are working in spherical polar coordinates, and $\overline{r} = r/R_M$
and $u = \cos \theta$. The function $\mu$ is given by
\begin{equation}
    \mu (\overline{r}, u ) = \frac{a_0}{g_N} \nu^\prime \left( \frac{g_N}{a_0} \right)
    \left[ \frac{2}{\overline{r}^7} - 4 \frac{\gamma_N}{a_0} \frac{u}{\overline{r}^5} + 
    \left( \frac{\gamma_N}{a_0} \right)^2
    \frac{(3 x^2 - 1)}{\overline{r}^3} \right]
    \label{eq:mu}
\end{equation}
and 
\begin{equation}
    \frac{g_N}{a_0} = \left[ \frac{1}{\overline{r}^4} + 
    \left( \frac{ \gamma_N }{a_0} \right)^2 - \frac{2}{\overline{r}^2} \frac{\gamma_N}{a_0} u  \right]^{1/2}.
    \label{eq:gnexplicit}
\end{equation}
Assuming the interpolating function is given by eq (\ref{eq:interpolfun}), the derivative is given by
\begin{equation}
    \nu^\prime (x) = - \frac{1}{2 \sqrt{x}} \frac{ \exp( - \sqrt{x} ) }{ [1 - \exp( - \sqrt{x} ) ]^2}.
    \label{eq:nuprime}
\end{equation}
Eqs (\ref{eq:phantomexplicitone}), (\ref{eq:mu}), (\ref{eq:gnexplicit}) and (\ref{eq:nuprime}) 
provide the desired explicit expression for $\rho_{{\rm ph}}$. 

To gain some insight into the expressions in the preceding paragraph we consider some limiting cases. 
First let us suppose that $\gamma_N = 0$. In the absence of the galactic field the expression for the
phantom mass simplifies to 
\begin{equation}
    \rho_{{\rm ph}} = \frac{M_\odot}{4 \pi R_M^3} \frac{1}{\overline{r}^4} 
    \frac{\exp( - 1/\overline{r} )}{ [ 1 - \exp( - 1/\overline{r} ) ]^2 }.
    \label{eq:phantomspherical}
\end{equation}
As expected the phantom mass distribution is spherically symmetric in the absence of the
galactic field.
Taking the limit $\overline{r} \rightarrow 0$ we find
\begin{equation}
\rho_{{\rm ph}} \rightarrow \frac{M_\odot}{4 \pi R_M^3 } \left( \frac{R_M}{r} \right)^4
\exp \left( - \frac{R_M}{r} \right);
\label{eq:sphericalphantomsmall}
\end{equation}
the phantom mass vanishes exponentially inside the MOND radius. For $\overline{r} \rightarrow 
\infty$ 
\begin{equation}
    \rho_{{\rm ph}} \rightarrow \frac{M_\odot}{4 \pi R_M^3} \left( \frac{R_M}{r} \right)^2;
    \label{eq:sphericalphantomlarger} 
\end{equation}
the phantom mass vanishes slowly as a power law far beyond the MOND radius.

Turning to the more relevant case where the galactic field is present
we find that for $\overline{r} \rightarrow 0$ the expression in eq (\ref{eq:sphericalphantomsmall})
still holds; the phantom mass vanishes exponentially within the solar system. For 
$\overline{r} \rightarrow \infty$ however the asymptotic behavior is different. 
In this limit
\begin{equation}
    \rho_{{\rm ph}} \rightarrow \frac{M_\odot}{4 \pi R_M^3} 
    \frac{ \sqrt{\eta_N} \exp ( - \sqrt{ \eta_N } ) }{[ 1 - \exp( - \sqrt{\eta_N} ) ]^2}
    \left( \frac{R_M}{r} \right)^3 P_2 (\cos \theta).
    \label{eq:phantomlargerasymptotics}
\end{equation}
Here for brevity we have written $\gamma_N / a_0 = \eta_N$.
Thus the phantom mass distribution vanishes more quickly as a power law when the
galactic field is taken into consideration. 
In fact it can be shown that the total phantom mass is finite so long as the external
galactic field is present \citep{quasimilgrom}. 

\begin{figure}
	\begin{center}
		\includegraphics[width=4.0in]{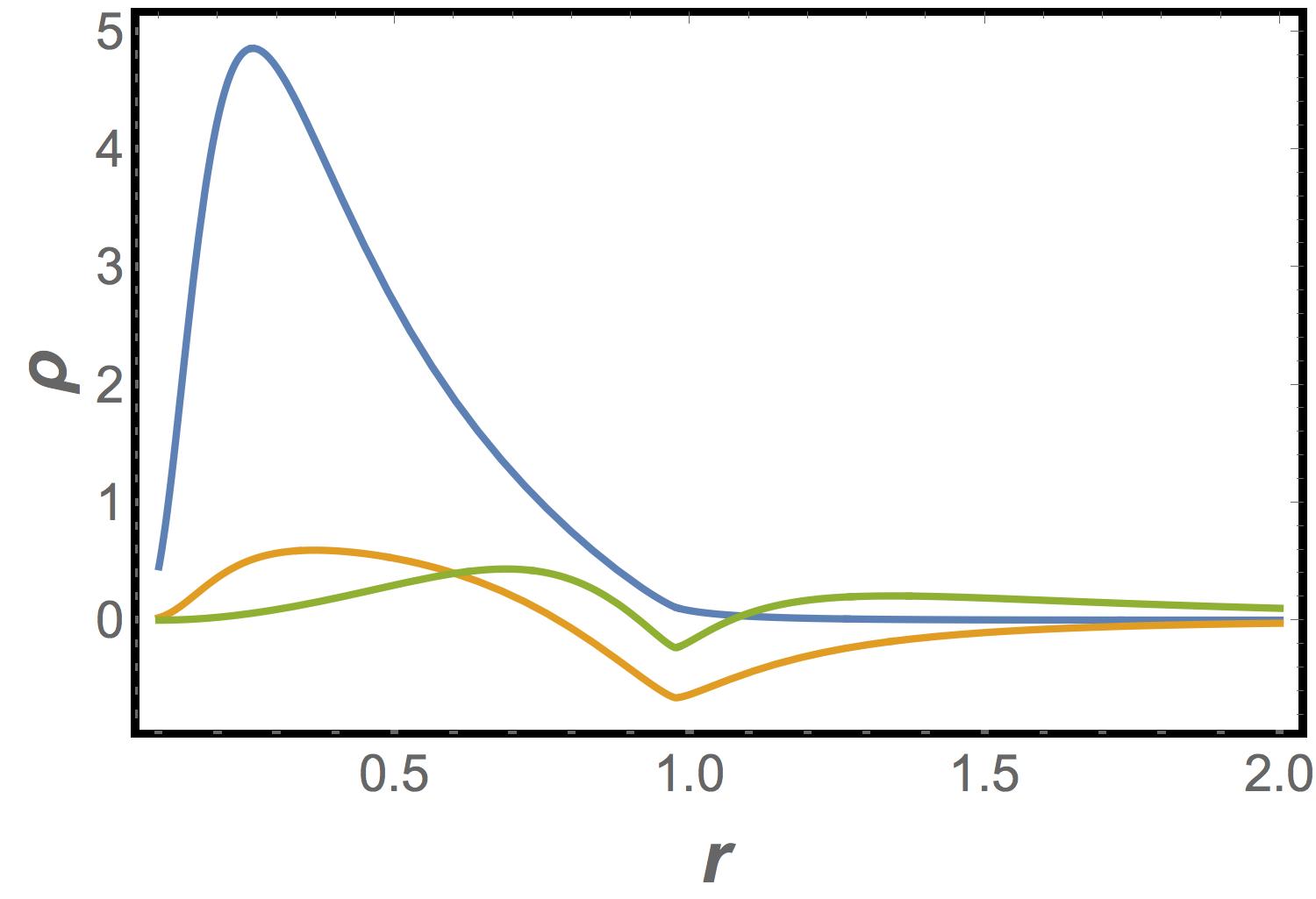}
		\caption{(Color online) Plot of the phantom density harmonics $\rho_{{\rm ph},\ell}$ as a
		function of distance from the sun for $\ell = 0$ (blue), $\ell = 1$ (green) and $\ell = 2$ (orange). 
		The density is in units of $M_\odot / R_M^3$ and 
		distance in units of $R_M$. In generating these plots we took
		$\gamma_g = 1.5 a_0$ as discussed in the main text. As expected the
		mass densities are peaked at a distance of order the MOND radius. The origin
		of the kinks at a common distance from the sun are discussed in the text.}
		\label{fig:harmonics}
	\end{center}
\end{figure}

The picture that emerges from the asymptotics is that there is essentially no 
phantom mass in the solar system well within the MOND radius. The phantom mass
density must be peaked at around the MOND radius and must decline gradually 
at greater distances. Moreover it is clear from the exact expression that
$\rho_{{\rm ph}}$ must be rotationally symmetric about the $z$-axis. 
We may therefore expand the
phantom mass distribution in harmonics defined via
\begin{equation}
    \rho_{{\rm ph} \ell} (\overline{r}) = \int_0^\pi d \theta \sin \theta P_{\ell} (\cos \theta) 
    \rho_{{\rm ph}} (\overline{r}, \theta). 
    \label{eq:phantomharmics} 
\end{equation}
These harmonics can be calculated numerically using the exact expression 
we have deduced for $\rho_{{\rm ph}}$ above. 
Fig \ref{fig:harmonics} shows the first few harmonics as a function of $r$. 
These plots confirm the general picture that $\rho_{{\rm ph}}$ is peaked 
near the MOND radius and exponentially small within it. The plots also
reveal a kink at a common value of $r$ in all harmonics. 

The origin of the kink is the following. There is a point along the
$z$ axis where the galactic field and the solar field exactly cancel. 
This point is known as a saddle point in the literature \citep{milgrom86, oria2021, 
penner2020}.
It is easy to see from eq (\ref{eq:newtonexternalfield}) that this 
distance is $\overline{r} = \sqrt{\eta_N}$. Near this point the
phantom mass diverges and moreover displays an interesting variation
in sign depending on the direction from which this point is approached.
However the density has a weak square root divergence at this point and
hence there is no substantial mass concentration present here. 

In order to calculate the harmonics numerically 
we have taken $\gamma_g$ the galactic gravitational field to be 
$1.5 a_0$ and solved the transcendental equation $\gamma_g = \gamma_N \nu (\gamma_N/a_0)$
to obtain $\gamma_N = 0.9 a_0$. The results are not sensitive to the particular value
of $\gamma_g/a_0$ so long as it is a factor of order unity. 
The value we have used is based on the rotation curve model
of the Milky Way described in \cite{stacymilkyway}. It is also compatible with
the value $\gamma_g = (1.9 \pm 0.1) a_0$ obtained from {\em Gaia} astrometry 
\citep{gaia2021}. 

\section{Multipole expansion in gravity}

The Newtonian gravitational potential $\psi_N$ is defined via the
relation ${\mathbf g}_N = - \nabla \psi_N$. For a mass distribution
$\rho$ the potential is given by the exact expression
\begin{equation}
    \psi_N ( {\mathbf r} ) = - \int d {\mathbf r}^\prime \frac{G \rho ({\mathbf r}^\prime)}{ | {\mathbf r} -
    {\mathbf r}' | }. 
    \label{eq:newtonpotexact}
\end{equation}
It is useful to recall the expansion \citep{matthews}
\begin{equation}
\frac{1}{| {\mathbf r} - {\mathbf r}^\prime | } = \sum_{\ell=0}^\infty \sum_{m = - \ell}^{+\ell} 
\frac{4 \pi}{(2 \ell + 1)} \frac{ r_<^\ell }{ r_>^{(\ell + 1)} } Y_{\ell m}^\ast (\theta^\prime, \phi^\prime)
Y_{\ell m} ( \theta, \phi ).
\label{eq:inverseexpansion}
\end{equation}
where $r_<$ denotes the lesser of $r$ and $r'$, and $r_>$ denotes the greater. 

For the circumstance that the mass distribution is localized near the origin, and the
observation point lies outside of the support of the mass distribution, by use of
eqs (\ref{eq:newtonpotexact}) and (\ref{eq:inverseexpansion}) we obtain 
\begin{equation}
    \psi_N ({\bf r}) = - \frac{GM}{R} \sum_{\ell = 0}^\infty \sum_{m = - \ell}^{\ell} J_{\ell m} 
    \left( \frac{R}{r} \right)^{\ell + 1} \sqrt{ \frac{4 \pi}{2 \ell + 1} } Y_{\ell m} 
    (\theta, \phi).
    \label{eq:conventionalmultipole}
\end{equation}
Here $M$ is the mass scale of the mass distribution, $R$ is the size scale 
and the dimensionless multipole moments $J_{\ell m}$ are given by
\begin{equation}
    J_{\ell m} = \frac{1}{M R^\ell} \int d {\mathbf r} \; \sqrt{ \frac{4 \pi}{2 \ell + 1} } r^\ell 
    Y^\ast_{\ell m} (\theta, \phi) \rho ({\mathbf r}).
    \label{eq:multipolemoments} 
\end{equation}
This is the usual multipole expansion familiar from electrostatics: it corresponds to the
situation that the mass distribution is localized and the field is desired at a point far from the mass 
distribution. As usual the $(\ell, m)$ term in the field falls off with distance as $1/r^{\ell +1}$
and has an angular dependence given by $Y_{\ell m}(\theta, \phi)$. 

Here we are interested in the complementary situation that within the mass distribution there
is an empty cavity devoid of mass. We are interested in the field near the center of this cavity.
Making use of eqs (\ref{eq:newtonpotexact}) and (\ref{eq:inverseexpansion}) we obtain
the complementary multipole expansion
\begin{equation}
    \psi_N ({\mathbf r}) = - \frac{GM}{R} \sum_{\ell=0}^\infty \sum_{-\ell}^\ell
    {\cal J}_{\ell m} 
    \left( \frac{r}{R} \right)^\ell
    \sqrt{ \frac{4 \pi}{2 \ell + 1} } Y_{\ell m} (\theta \phi).
    \label{eq:complementarymultipole}
\end{equation}
Again $M$ is the mass scale of the mass distribution and $R$ the size scale. 
The dimensionless complementary multipole moments ${\cal J}_{\ell m}$
are given by
\begin{equation}
    {\cal J}_{\ell m} = \frac{R^{\ell +1}}{M} \sqrt{\frac{4 \pi}{2 \ell + 1}} 
    \int d {\mathbf r} \frac{1}{r^{\ell + 1}} Y_{\ell m}^\ast (\theta, \phi) \rho ({\mathbf r}). 
    \label{eq:complementarymoments}
\end{equation}
In contrast to the usual multipole expansion we see that here the $(\ell, m)$
term in the expansion grows as $r^\ell$; its angular dependence is still given by
$Y_{\ell m} (\theta, \phi)$.

For the special case of cylindrical symmetry ${\cal J}_{\ell m} = 0$
except for $m = 0$. Hence the complementary multipole expansion simplifies to
\begin{equation}
    \psi_N ({\mathbf r}) = - \frac{GM}{R} \sum_{\ell = 0}^\infty {\cal J}_{\ell 0} 
    \left( \frac{r}{R} \right)^\ell P_\ell (\cos \theta).
    \label{eq:compmultipolecylinder}
\end{equation}
and the non-vanishing multipole moments are given by
\begin{equation}
    {\cal J}_{\ell 0} = \frac{R^{\ell + 1}}{M} \int d {\mathbf r} \frac{1}{r^{\ell +1}} P_\ell (\cos \theta)
    \rho ({\mathbf r}).
    \label{eq:complcylmoments}
\end{equation}

For reference it is useful to explicitly write the quadrupole term for the case of
cylindrical symmetry. 
\begin{equation}
    \psi^Q_N ({\mathbf r}) = - \frac{GM}{R^3} {\cal J}_{2 0} r^2 P_2 (\cos \theta)
    \label{eq:quadcylcomp}
\end{equation}
where
\begin{equation}
    {\cal J}_{2 0} = \frac{R^3}{M} \int d {\mathbf r} \frac{1}{r^3} P_2 (\cos \theta) \rho ({\mathbf r}).
    \label{eq:quadcylcompmoment}
\end{equation}

Thus far in our discussion we have focused on the multipole expansion for the
Newtonian gravitational field ${\mathbf g}_N = - \nabla \psi_N$. However the 
same discussion applies, mutatis mutandis, for the MOND external field
${\mathbf g}_A = - \nabla \psi_A$ which is mathematically the Newtonian
field produced by the phantom mass $\rho_{{\rm ph}}$. 

We now evaluate the quadrupole term in the multipole expansion
(\ref{eq:complementarymultipole}) for several cases of interest.
The simplest mass distribution that produces a quadrupolar field
is two point masses each of mass $M$ located on the $z$-axis
symmetrically about the origin at a distance $2R$ from each other.
This distribution has cylindrical symmetry and a simple calculation
using eq (\ref{eq:quadcylcompmoment}) 
reveals that ${\cal J}_{20} = 2$ in this case. Note that the multipole
moment is positive here. This is called a prolate quadrupole.

Next consider a circle of radius $R$ and total mass $M$ that lies in the
$x$-$y$ plane centered about the origin. This distribution also has
cylindrical symmetry and in this case use of eq (\ref{eq:quadcylcompmoment}) 
reveals that ${\cal J}_{20} = - \frac{1}{2}$ here. Note that the 
multipole moment is negative here. This is called an oblate quadrupole. 

Next let us consider the quadrupolar field of a hypothetical planet nine. 
In the secular approximation we want the orbit averaged field of planet nine.
In other words we assume that the mass of planet nine is distributed non-uniformly
over the ellipse corresponding to its orbit. The amount of mass contained in any
arc of the orbit is proportional to the time it takes planet nine to traverse
that arc. We assume that the orbit lies in the $x$-$y$ plane with the sun
at the origin and the point of perihelion on the positive $x$-axis. 
Note that this mass distribution does not have cylindrical symmetry.
Nonetheless we find upon explicit evaluation that ${\cal J}_{2 m} = 0$
except for $m = 0$. For that case, setting $M \rightarrow m_9$ and 
$R \rightarrow a_9$, where $m_9$ is the mass of planet nine and $a_9$ is the 
semimajor axis of its orbit, we obtain
\begin{equation} 
{\cal J}_{20} = - \frac{1}{2} \frac{1}{(1 - e_9^2)^{3/2}}
\label{eq:planetninequadmomentcomp}
\end{equation}
where $e_9$ is the eccentricity of the orbit of planet nine. 
Note that eq (\ref{eq:planetninequadmomentcomp}) reduces to the 
expected result for a circle when $e_9 = 0$. 
It follows from eq (\ref{eq:complementarymultipole})
that the quadrupolar field of planet nine is cylindrically symmetric 
and given by
\begin{equation}
    \psi^Q_9 ({\mathbf r}) = 
    \frac{1}{2} \frac{G m_9}{a_9^3} \frac{1}{(1 - e_9^2)^{3/2} } r^2 P_2 (\cos \theta)
    \label{eq:planetninequadrupole}
\end{equation}
We postpone for now the derivation of the multipole moments 
${\cal J}_{2m}$ for the orbit of planet nine quoted above
but we note that the emergent cylindrical 
symmetry in the moments is analogous to the well known fact
that the moment of inertia of a cube is isotropic and diagonal. A cube
is by no means spherically symmetrical but it has sufficient
symmetry to render the moment of inertia isotropic. 

We note that the dominant interaction of planet nine with
objects in the inner solar system is via its quadrupole field.
The reader may have noticed that the 
orbit averaged mass distribution has a non-zero
dipole moment, but it is well known that in an inertial frame
centered at the sun the dipole interaction term is 
exactly cancelled by the ``indirect interaction'' potential
(see for example section 6-2 of \cite{elliotplusmcdermott}). 

We turn now to the multipole expansion of the anomalous field 
in MOND. The phantom mass distribution is cylindrically symmetric
about the z-axis, which is taken to point towards the galactic
center, with the sun at the origin. The non-vanishing multipole
moments are written as ${\cal J}_{\ell 0} = q_\ell / 4 \pi$ where
\begin{equation}
    q_\ell = - 2 \pi \int_0^\infty d \overline{r} \int_{-1}^{+1} d u 
    \overline{r}^{1 - \ell} P_\ell (u) \mu (\overline{r}, u)
    \label{eq:phantomultipole}
\end{equation}
where $\mu(\overline{r}, u)$ is given by eq (\ref{eq:mu}).
If we choose the galactic field $\gamma_g/a_0 = 1.5$
as in the previous section then we find by numerical evaluation of 
eq (\ref{eq:phantomultipole}) the following values
for the lowest order multipoles
\begin{equation}
    q_1 \approx 0, \hspace{3mm}
    q_2 = 1.00, \hspace{3mm}
    q_3 = 0.51 \ldots
    \label{eq:lowestmultipoles}
\end{equation}
That $q_1 \approx 10^{-12}$ in our numerical evaluation is 
encouraging since it is known that $q_1 = 0$ exactly
for the phantom mass. This is shown by
\cite{milgromgalacticefe} and \cite{milgromefe}.
Thus for MOND, as for planet nine, its dominant ineraction
with bodies in the inner solar system is quadrupolar. 
We note that the values of $q_2$ and $q_3$ are functions of
the dimensionless quantity $\gamma_g/a_0$ and are also sensitive
to the form of the interpolating function (this sensitivity is
studied by \cite{milgromefe}). Thus at the present stage
of development MOND is not able to pin down the precise values
of these parameters. The takeaway from our computation and the
tabulation by \cite{milgromefe} is that 
$q_2$ and $q_3$ are of order unity to within an order of magnitude 
or two. Studies of the outer solar system will help constrain
the values of these parameters and hence the interpolating 
function in a way that is independent of and complementary
to traditional studies of galaxy rotation \citep{rar}.
The parameter $q_2$ is also bounded by spacecraft orbits.
The best bounds are from Cassini \citep{hees2014}. 
Our parameter $q_2$ is related to the
parameter $Q_2$ in \citep{hees2014} via 
\begin{equation}
q_2 = \frac{4 \pi}{3} \sqrt{ \frac{ G M_{{\rm sun}} }{ a_0^3} } Q_2. 
\label{eq:heescopy}
\end{equation}
$q_2 = 1.00$ corresponds to $Q_2 = 27 \times 10^{-27}$ s$^{-2}$ which is in tension
with the Cassini bound of $(3 \pm 3) \times 10^{-27}$ s$^{-2}$ \citep{hees2014}.
However, \cite{hees2016} have shown that other interpolating functions yield
values of $Q_2$ that are compatible with the Cassini bound. Our results
on orbital dynamics are not sensitive to the form of the interpolating function
and a quadrupole compatible with the Cassini bound is sufficient for our
purpose. 

For reference we now provide an explicit 
expression for the quadrupole field in MOND.
\begin{equation}
    \psi_A^Q = - \frac{G M_\odot}{R_M^3} \frac{q_2}{4 \pi} r^2 P_2 (\cos \theta).
    \label{eq:mondquadrupole} 
\end{equation}
Eq (\ref{eq:planetninequadrupole}) and eq (\ref{eq:mondquadrupole}) 
provide expressions for the quadrupolar fields that are used in the
paper to compare the order of magnitude of the MOND perturbation with
that of planet nine on objects in the Kuiper belt. Note that
the MOND potential has a spurious reflection symmetry
about the $x$-$y$ plane in the quadrupole approximation.
This symmetry is absent in the problem since there is 
difference between the direction pointing towards
galactic center and the direction pointing away from it. 
The octupole term breaks this symmetry and inclusion 
of this term will prove important for our analysis below. 
The multipole expansion is a good approximation for $r \ll R_M = 7000$ au. 
For a Kuiper belt object whose orbit extends to 500 au this 
condition is reasonably well satisfied.

\begin{figure}
	\begin{center}
		\includegraphics[width=4.0in]{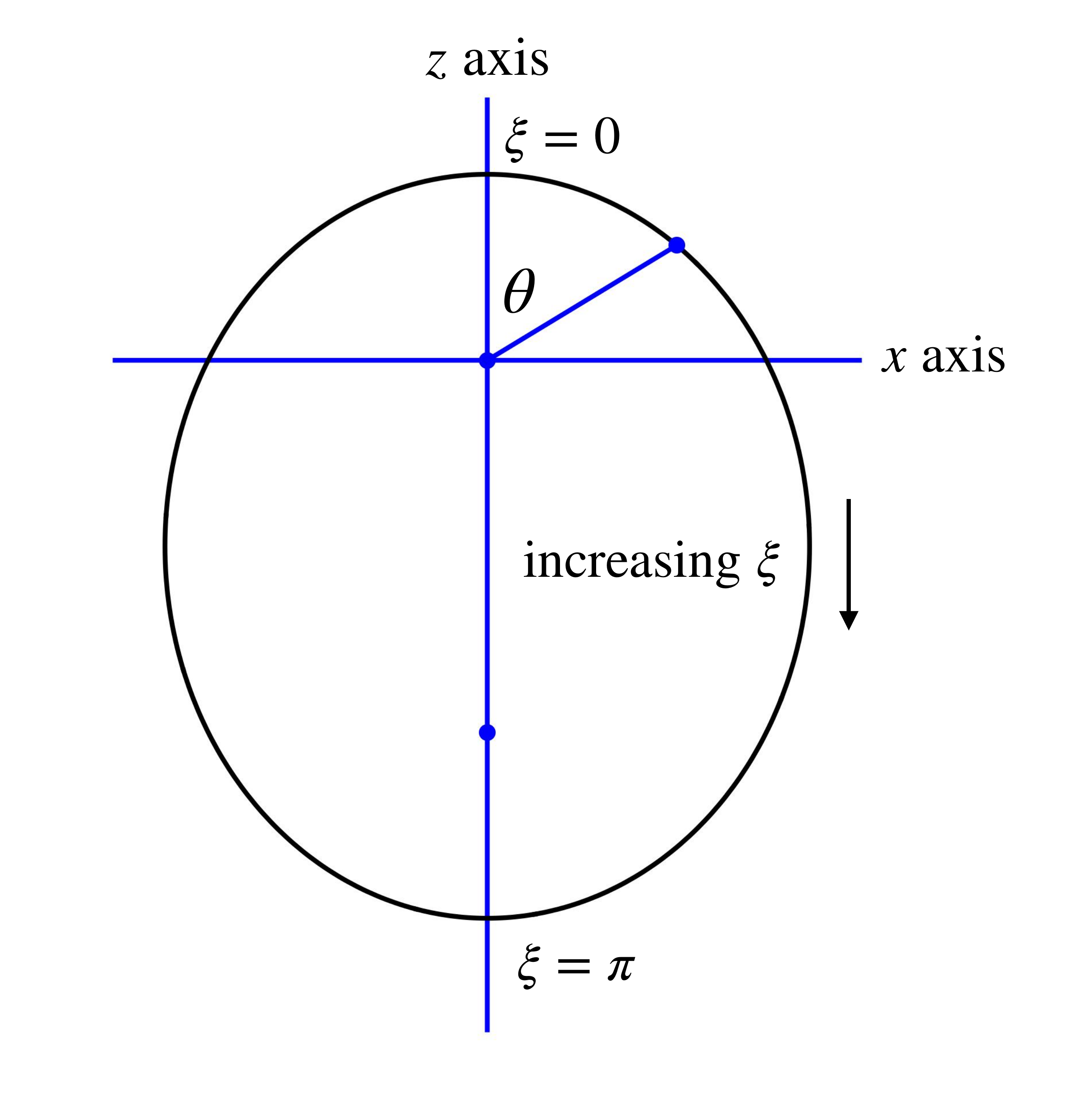}
		\caption{(Color online) The orbit of planet nine is taken to lie in the $z$-$x$ plane
		with the sun at the origin. The perihelion is assumed to lie on the positive $z$-axis.
		The multipole moments for this reference configuration are denoted $J^{(0)}$ and 
		${\cal J}^{(0)}$. The eccentric anomaly $\xi$ is taken to be zero at perihelion and 
		to increase clockwise in the figure. Also shown in the figure is the 
		angle $\theta$ for an arbitrary point on the orbit. $\theta$ is the spherical 
		polar colatitude of that point and it also represents the true anomaly 
		around the orbit.}
		\label{fig:kbofig}
	\end{center}
\end{figure}

For later use let us also calculate the orbit averaged multipole
moments for planet nine. We take the orbit to lie in the
$z$-$x$ plane with the sun at the origin and the perihelion 
along the positive $z$-axis. The ellipse may then be described
parametrically by the equations
\begin{equation}
    z = a_9 ( \cos \xi - e_9 ) \hspace{3mm} {\rm and} \hspace{3mm}
    x = a_9 \sqrt{ 1 - e_9^2 } \sin \xi.
    \label{eq:parametericellipse}
\end{equation}
Here the parameter $\xi$ is known as the eccentric anomaly and
ranges from $0$ to $2 \pi$ around the ellipse (see fig \ref{fig:kbofig}). It is useful
to also calculate the spherical polar coordinates $(r, \theta, \phi)$
of points around the ellipse as a function of $\xi$. It follows from
eq (\ref{eq:parametericellipse}) that
\begin{equation}
    r = a_9 (1 - e_9 \cos \xi)
    \label{eq:rofxi}
\end{equation}
and that 
\begin{equation}
    \cos \theta = \frac{ \cos \xi - e_9 }{1 - e_9 \cos \xi } 
    \hspace{3mm} {\rm and} \hspace{3mm}
    \sin \theta = \frac{ \sqrt{1 - e_9^2} \sin \xi }{1 - e_9 \cos \xi}.
    \label{eq:thetaofxi}
\end{equation}
Finally $\phi = 0$ for $0 < \xi < \pi$ and $\phi = \pi$ for
$\pi < \xi < 2 \pi$. 
Let us assume that the planet starts at $\xi = 0$ at $t = 0$
and moves in the direction of increasing $\xi$. The anomaly
$\xi$ of the planet at time $t$ is then given by Kepler's 
anomaly equation
\begin{equation} 
\frac{2 \pi t}{T_9} = \xi - e_9 \sin \xi \hspace{2mm}
\Leftrightarrow \hspace{2mm} \frac{2 \pi}{T_9} \frac{d t}{d \xi} = 1 - e_9 \cos \xi. 
\label{eq:anomalyeq}
\end{equation}
Here $T_9$ denotes the orbital period of planet nine. 
The mass $d m$ associated with an infinitesimal arc of the
orbit is proportional to the time it takes to traverse the
arc. Hence $d m / m_9 = d t / T_9$ and making use of eq (\ref{eq:anomalyeq})
we then obtain
\begin{equation}
    \frac{d m}{d \xi} = \frac{m_9}{2 \pi} (1 - e_9 \cos \xi)
    \label{eq:orbitmassdistn} 
\end{equation}

We are now in a position to compute the orbit averaged 
multipole moments. We find that for even $m$
\begin{equation}
    J^{(0)}_{\ell m} = \frac{1}{\pi} \sqrt{ \frac{ (\ell - m )! }{ ( \ell + m )! } }
    \int_0^\pi d \xi (1 - e_9 \cos \xi)^{\ell + 1} 
    P_{\ell, m} \left( \frac{ \cos \xi - e_9 }{1 - e_9 \cos \xi} \right).
    \label{eq:innermultipoles}
\end{equation}
and for odd $m$ 
$J^{(0)}_{\ell m} = 0$.
The superscript in $J^{(0)}$ is meant to remind us that this multipole
moment is computed with the orbit in the $z$-$x$ plane with the sun
at the origin and the perihelion along the positive $z$-axis. 
Eq (\ref{eq:innermultipoles}) was obtained from eq (\ref{eq:multipolemoments})
by making the following substitutions. 
$M \rightarrow m_9$, $R \rightarrow a_9$, and 
$d {\mathbf r} \rho \rightarrow d \xi ( d m_9 / d \xi)$, using
eq (\ref{eq:orbitmassdistn}) for $dm/d \xi$, 
eq (\ref{eq:rofxi}) for $r$ and eq (\ref{eq:thetaofxi}) for 
$\cos \theta$. It is also helpful to recall that 
\begin{equation}
    Y_{\ell, m} (\theta, \phi) = \sqrt{ \frac{2 \ell + 1}{4 \pi} } 
    \sqrt{ \frac{ (\ell - m)! }{ (\ell + m )! } } P_{\ell m} ( \cos \theta ) e^{ i m \phi }.
    \label{eq:sphericalharmonic}
\end{equation}
The integral in eq (\ref{eq:innermultipoles}) is evaluated only over
the right half of the orbit ($0 < \xi < \pi$) for which 
$e^{i m \phi} = 1$. For the left half ($\pi < \xi < 2 \pi$) the
quantity $e^{i m \phi} = (-1)^m$. Hence the contribution of the right
half must be doubled for even $m$ and for odd $m$ the two halves cancel
leading to $J^{(0)}_{\ell m} = 0$. 

For the record we explicitly provide the low order multipoles
that are most relevant to our analysis. We find upon integrating
eq (\ref{eq:innermultipoles}) that $J^{(0)}_{0,0} = 1$. 
The non-zero quadrupole moments are found to be
\begin{equation}
    J^{(0)}_{2, \pm 2} = \frac{1}{4} \sqrt{ \frac{3}{2} } (1 - e_9^2)
    \hspace{3mm} {\rm and} \hspace{3mm} J^{(0)}_{2, 0} = \frac{1}{4} ( 1 + 9 e_9^2 ).
    \label{eq:innerquadrupolemoments}
\end{equation}
The non-zero octupole moments are found to be
\begin{equation}
    J^{(0)}_{3, \pm 2} = - \frac{5}{16} \sqrt{\frac{15}{2}} e_9 (1 - e_9^2 )
    \hspace{3mm} {\rm and} \hspace{3mm} 
    J^{(0)}_{3, 0} = - \frac{5}{16} e_9 (3 + 11 e_9^2).
    \label{eq:inneroctupolemoments} 
\end{equation}
Obviously the same expressions would apply for a Kuiper belt object with the
replacement $e_9 \rightarrow e_K$ where $e_K$ is the eccentricity of the Kuiper
belt object or for Neptune with the replacement $e_9 \rightarrow e_8$ where 
$e_8$ is the eccentricity of the orbit of Neptune. 

The complementary multipole moments ${\cal J}^{(0)}_{\ell m}$ may be
calculated similarly using eq (\ref{eq:complementarymoments}). 
Analogous to eq (\ref{eq:innermultipoles}) we obtain
\begin{equation}
    {\cal J}^{(0)}_{\ell m} = \frac{1}{\pi} \sqrt{ \frac{ (\ell - m)! }{ (\ell + m)! } } 
    \int_0^\pi d \xi \frac{1}{(1 - e_9 \cos \xi)^\ell} P_{\ell m} 
    \left( \frac{\cos \xi - e_9}{1 - e_9 \cos \xi}  \right).
    \label{eq:outermultipoles}
\end{equation}
For later use we provide low order complementary multipole moments that
will be especially relevant. By explicit evaluation of the integral in
eq (\ref{eq:outermultipoles}) we find ${\cal J}^{(0)}_{00} = 1$. 
The complementary quadrupole moments are 
\begin{equation}
    {\cal J}^{(0)}_{2, \pm 2} = \frac{1}{4} \sqrt{ \frac{3}{2} } \frac{1}{(1 - e_9^2)^{3/2}}
    \hspace{3mm} {\rm and} \hspace{3mm} 
    {\cal J}^{(0)}_{2,0} = \frac{1}{4} \frac{1}{(1 - e_9^2)^{3/2}}. 
    \label{eq:planetnineouterquadrupole}    
\end{equation}
Note that eq (\ref{eq:planetnineouterquadrupole}) does not match eq (\ref{eq:planetninequadmomentcomp}) 
because they are computed for different orientations of the orbit. We describe in the
next section how to obtain eq (\ref{eq:planetninequadmomentcomp}) from 
eq (\ref{eq:planetnineouterquadrupole}).

\section{The Disturbing Function}

As a prelude consider two distributions of mass: $\rho_1$ localized near the origin
and $\rho_2$ located far from the origin. The gravitational potential energy
of these two distributions is
\begin{equation} 
{\cal R} = - G \int d {\mathbf r} \int d {\mathbf r}^\prime \;
\frac{\rho_1 ({\mathbf r}^\prime) \rho_2 ({\mathbf r})}{ | {\mathbf r} - {\mathbf r}^\prime | }
= \int d {\mathbf r} \psi_1({\mathbf r}) \rho_2 ({\mathbf r}).
\label{eq:gravpot}
\end{equation}
Here $\psi_1$ denotes the gravitational potential due to $\rho_1$. 
Making use of the multipole expansion eq (\ref{eq:conventionalmultipole}) and
the definition of complementary moments eq (\ref{eq:complementarymoments}) 
we obtain 
\begin{equation}
    {\cal R} = - \frac{G M_1 M_2}{R_2} \sum_{\ell = 0}^\infty \sum_{m=-\ell}^\ell 
    \left( \frac{ R_1 }{ R_2 } \right)^\ell {\cal J}^\ast_{\ell m} J_{\ell m}.
    \label{eq:gravpotmultipole}
\end{equation}
Here $M_1$ is the mass scale for the distribution $\rho_1$ and $R_1$ the 
length scale. $M_2$ and $R_2$ are the corresponding quantities for $\rho_2$. 
If we regard $J_{\ell m}$ as the $2 \ell + 1$ components of a column vector 
$J_\ell$ and ${\cal J}_{\ell m}^\ast$ as the components of a row vector
${\cal J}_\ell^\dagger$ we may write eq (\ref{eq:gravpotmultipole}) more
compactly as 
\begin{equation}
    {\cal R} = - \frac{G M_1 M_2}{R_2} \sum_{\ell = 0}^\infty 
    \left( \frac{ R_1 }{R_2} \right)^\ell 
    {\cal J}^\dagger_\ell J_\ell. 
    \label{eq:gravpotmultipolematrixform}
\end{equation}

Now let us consider the effect of the MOND field ${\mathbf g}_A$ 
on the motion of a Kuiper belt object (KBO). Under the influence
of the sun alone the KBO would trace an ellipse in accordance
with Kepler's laws. The ellipse is characterized by five 
orbital elements $(a_K, e_K, \omega_K, i_K, \Omega_K)$. 
The semimajor axis $a_K$ and the eccentricity $e_K$ specify
the shape of the orbit. The Euler angles $(\omega_K, i_K, \Omega_K)$ 
specify the orientation and will be defined more precisely momentarily. 
Under the influence of perturbations the orbital elements change slowly.
In the secular approximation the semimajor axis does not change and
the dynamics is controlled by the disturbing function ${\cal R}$
which is the potential energy of the KBO time averaged over its Keplerian orbit
in the perturbing MOND potential \citep{elliotplusmcdermott, doverbook}.

The disturbing function is often calculated in an expansion in the
eccentricity of the orbit or the inclination of the orbit \citep{elliotplusmcdermott}.
In our problem we need the disturbing function for arbitrary eccentricity
and inclination and hence we cannot use expansions in these variables.
We have developed a method to calculate the secular disturbing function 
by expanding it in a multipole expansion and computing the quadrupole
and octupole terms exactly. In principle the higher multipoles can also
be calculated by this method but they are not needed for our immediate
purpose.

The key ingredient in our approach is to recognize that the 
multipole moments $J_\ell$ and ${\cal J}_\ell$ transform as rank $\ell$
tensors under rotations. Thus if, for example, a mass distribution $\rho$ is rotated
about the $y$-axis by an angle $\varphi$ then the multipole moments transform
as $J_\ell \rightarrow R^{(\ell)}_y (\varphi) J_\ell$ and similarly for
${\cal J}_\ell$. Here $R^{(\ell)}_y (\varphi)$ is a $(2 \ell + 1) \times (2 \ell + 1) $
dimensional unitary matrix that corresponds to the rotation about the $y$-axis for a 
rank $\ell$ tensor. We will discuss the form of these matrices shortly
but first let us use them to develop an expression for the disturbing
function. 

In this section we will find it convenient to 
work in a coordinate frame wherein the sun is at the origin and the
$y$-axis points in the direction of the galactic center. The multipole 
moments of the phantom mass distribution are then given by
\begin{equation}
    {\cal J}_\ell = R^{(\ell)}_x \left( - \frac{\pi}{2} \right) {\cal J}^{(0)}_\ell 
    \label{eq:phantomj}
\end{equation}
where $R^{(\ell)}_x ( - \pi / 2 )$ denotes the matrix corresponding to a
rotation about the $x$-axis by $-\pi/2$ and 
\begin{equation}
    {\cal J}^{(0)}_{\ell m} = \frac{q_\ell}{4 \pi} \delta_{m,0}
    \label{eq:jzerophantom}
\end{equation}
are the multipole moments for the phantom mass distribution when the galactic center
is taken to be along the $z$-axis as it was in the preceding section. 
$q_\ell$ is given by eqs (\ref{eq:phantomultipole}) and (\ref{eq:lowestmultipoles}). 
We will compute the components of ${\cal J}_\ell$ below explicitly for $\ell =2$
and $\ell = 3$. 

The $z$-$x$ plane will be used 
as the reference plane to describe the evolution of the
KBO orbit. We take as the reference orientation the configuration such that the
KBO orbit lies in the reference plane with its perihelion located on the positive
$z$-axis. The orbital elements $(\omega_K, i_K, \Omega_K)$ correspond to the 
orientation obtained by first rotating the reference orbit about the $y$-axis by the
angle $\omega_K$, then about the $z$-axis by the angle $i_K$ and finally about
the $y$-axis again by an angle $\Omega_K$. The orbit orientation is also fully
specified by two unit vectors: the apsidal vector $\hat{{\mathbf a}}_K$
which points from the sun to the direction of the perihelion and the orbit normal
$\hat{{\bf n}}_K$ which is normal to the plane of the orbit and points in the
direction of the angular momentum of the KBO. In the reference configuration
$\hat{{\bf a}}^{(0)}_K = \hat{{\bf z}}$ and $\hat{{\bf n}}^{(0)}_K = \hat{{\bf y}}$. 
The reader may find it helpful to visualize how $\hat{{\bf a}}_K$ and $\hat{{\bf n}}_K$
change as the orbit undergoes successive $y$-$z$-$y$ rotations by the angles
$(\omega_K, i_K, \Omega_K)$. Note in particular that $\hat{{\mathbf n}}_K$ will be
inclined at an angle $i_K$ to the positive $y$-axis at the end of the process. 
The orbit averaged multipole moments of the KBO with orbital elements 
$(a_K, e_K, \omega_K, i_K, \Omega_K)$ is given by
\begin{equation}
    J_\ell = R^{(\ell)}_y (\Omega) R^{(\ell)}_z (i_K) R^{(\ell)}_y (\omega_K) J_\ell^{(0)}
    \label{eq:kboj}
\end{equation}
where $J_\ell^{(0)}$ denotes the moments for the standard orientation of the
orbit that we calculated in the preceding section. 
(See eqs (\ref{eq:innermultipoles}), (\ref{eq:innerquadrupolemoments}) and 
(\ref{eq:inneroctupolemoments}) with the replacement 
$e_9 \rightarrow e_K)$.

We can now write an exact formal expression for the disturbing function that
governs the evolution of the KBO orbit under the influence of the MOND field. 
We substitute in eq (\ref{eq:gravpotmultipolematrixform}) the expression for
${\cal J}_\ell$ given in eq (\ref{eq:phantomj}) and the expression for
${\cal J}_\ell$ given in eq (\ref{eq:kboj}) to obtain
\begin{equation}
    {\cal R} = - \frac{G m_K M_\odot}{R_M} \sum_{\ell = 2}^\infty 
    \left( \frac{a_K}{R_M} \right)^\ell {\cal M}_\ell
    \label{eq:mondisturb}
\end{equation}
where the matrix element ${\cal M}_\ell$ is given by
\begin{equation}
{\cal M}_\ell = {\cal J}_\ell^\dagger R^{(\ell)}_y (\Omega_K) R^{(\ell)}_z (i_K) R^{(\ell)}_y (\omega_K) 
    J^{(0)}_\ell. 
    \label{eq:prematrixelement}
\end{equation}
${\cal M}_\ell$ depends on $e_K$ through $J^{(0)}_\ell$
and on $(\omega_K, i_K, \Omega_K)$ through the rotation factors. It also depends on
$q_\ell$ through ${\cal J}_\ell$. 
Note that we have omitted the $\ell = 0$ term because it is independent of all the
orbital elements and therefore does not have any effect on the dynamics. We are able to omit the
$\ell = 1$ term because the phantom mass has no dipole moment. Intuitively we 
expect that due to the cylindrical symmetry of the MOND field about the $y$-axis,
the matrix element ${\cal M}$ and the disturbing function
should be independent of $\Omega_K$. Mathematically this
happens because ${\cal J}_\ell$ in eq (\ref{eq:phantomj}) proves to be 
invariant under rotations about the $y$-axis. Thus $R^{(\ell)}_y (\varphi) {\cal J}_\ell =
{\cal J}_\ell$ for any angle $\varphi$, 
which implies ${\cal J}_{\ell}^\dagger R^{(\ell)}_y (\Omega_K) = 
{\cal J}^{\dagger}_\ell$.

We may write the rotation matrix $R^{(\ell)}_y (\varphi) = \exp [ - i L_y^{(\ell)} \varphi ]$
where $L_y^{(\ell)}$ is the $(2 \ell + 1) \times (2 \ell + 1)$ matrix
corresponding to the generator of rotations about the $y$-axis. From the general
theory of representations of the rotation group (familiar to physicists as the
quantum theory of angular momentum) we know that $L^{(\ell)}_y$ has 
$2 \ell + 1$ eigenvectors $v^{(y)}_{\ell \mu}$ with eigenvalue $\mu$ 
where $\mu$ is an integer in the range $- \ell \leq \mu \leq \ell$.
In terms of these eigenvectors $R_y^{(\ell)} (\varphi)$ has the spectral 
representation 
\begin{equation}
    R_y^{(\ell)} (\varphi) = \sum_{\mu = - \ell}^{\ell} \exp ( - i \mu \varphi) 
    v^{(y)}_{\ell \mu} v^{(y) \dagger}_{\ell \mu}
        \label{eq:spectraly} 
\end{equation}
Note that $v_{\ell \mu}^{(y)}$ is a $2 \ell + 1$ component column vector;
$v_{\ell \mu}^{(y) \dagger}$ is a $2 \ell + 1$ component row vector. 
Hence the right hand side of eq (\ref{eq:spectraly}) is a $(2 \ell + 1) \times (2 \ell + 1)$
matrix as it should be. Sometimes we will find it convenient to explicitly write the
components of $v_{\ell \mu}^{(y)}$ as $v_{\ell \mu}^{(y)} (m)$ where $m$ is an 
integer in the range $- \ell \leq m \leq \ell$.
Similarly we may write the spectral representation
\begin{equation}
    R_x^{(\ell)} (\varphi) = \sum_{\mu = - \ell}^{\ell} \exp( - i \mu \varphi ) v^{(x)}_{\ell \mu} 
    v^{(x) \dagger}_{\ell \mu} 
    \label{eq:spectralx}
\end{equation}
where $v^{(x)}_{\ell \mu}$ are eigenvectors of $L^{(\ell)}_x$, the generator of rotations
about the $x$-axis. 
We are working in a basis wherein $L^{(\ell)}_z$ is diagonal. 
Hence the eigenvectors are $v^{(z)}_{\ell \mu} (m) = \delta_{\mu,m}$
and the matrix $R_z^{(\ell)} (\varphi)$ is diagonal. Its effect on a vector is
simply to multiply each component of the vector by the appropriate phase. 
Thus
\begin{equation}
    R_z^{(\ell)} (\varphi) J_\ell \rightarrow \exp( - i m \varphi ) J_{\ell m}.
    \label{eq:rzmatrix}
\end{equation}

With the help of these results we arrive at the following expression for the
matrix element ${\cal M}_\ell$
in eq (\ref{eq:matrixelement}). 
\begin{equation} 
  {\cal M}_\ell = \sum_{\mu \mu'} e^{- i \mu \Omega_K}
    \left[ {\cal J}_\ell^\dagger v^{(y)}_{\ell \mu} \right] {\cal I}_{\mu \mu^\prime} 
    \left[ v^{(y) \dagger}_{\ell \mu^\prime} J_\ell^{(0)} \right] 
    e^{- i \mu^\prime \omega_K}
        \label{eq:matrixelement}
\end{equation}
where the sums over $\mu$ and $\mu^\prime$ range from $-\ell$ to $\ell$ and
the inclination factor ${\cal I}$ is given by
\begin{eqnarray}
    {\cal I}_{\mu \mu^\prime} & = & 
    \sum_{\mu^{\prime \prime} = \ell}^\ell 
    \left[ v^{(y) \dagger}_{\ell \mu} v^{(z)}_{\ell \mu^{\prime \prime}} \right]
    \exp ( - i \mu^{\prime \prime} i_K ) 
    \left[ v^{(z) \dagger}_{\ell \mu^{\prime \prime}} v^{(y)}_{\ell \mu^\prime} \right] 
    \nonumber \\
    & = &
    \sum_{\mu^{\prime \prime} = - \ell}^{\ell} 
    v^{(y)}_{\ell \mu} (\mu^{\prime \prime}) \exp ( - i \mu^{\prime \prime} i_K ) 
    v^{(y)}_{\ell \mu^\prime} ( \mu^{\prime \prime} ).
    \label{eq:inclinationfactor} 
\end{eqnarray}
Eqs (\ref{eq:matrixelement}) and (\ref{eq:inclinationfactor}) provide the tools
needed to calculate the matrix element ${\cal M}_\ell$ and hence
disturbing function. We illustrate the computation for
$\ell = 2$ and also carry it out explicitly for $\ell = 3$ since we need the 
disturbing function to octupole order for our analysis. However we note that
closed form expressions for $v^{(y)}_{\ell \mu}$, ${\cal J}_\ell$ and 
$J^{(0)}_\ell$ exist for arbitrary $\ell$ and hence the procedure outlined
above allows us to compute the matrix element 
for any $\ell$ 
exactly without perturbing in $e_K$ or $i_K$. 

To obtain ${\cal M}_2$ our first task is to compute ${\cal J}_2$ for
the phantom mass distribution. Using the spectral representation for
$R_x ( - \pi / 2)$ we may write eq (\ref{eq:phantomj}) as
\begin{equation}
    {\cal J}_2 = \sum_{\mu = -2}^2 v^{(x)}_{2 \mu} 
    \left[ v^{(x) \dagger}_{2 \mu} {\cal J}_2^{(0)} \right].
    \label{eq:phantomj2computed}
\end{equation}
Making use of the explicit formulae for $v^{(x)}_{2 \mu}$ tabulated below and
eq (\ref{eq:jzerophantom}) for ${\cal J}_2^{(0)}$ we obtain 
\begin{equation}
    v^{(x) \dagger}_{2, \pm 2} {\cal J}_2^{(0)} = \frac{q_2}{4 \pi} \frac{ \sqrt{6} }{4} 
    \hspace{2mm} {\rm and} \hspace{2mm}
    v^{(x) \dagger}_{2, 0} {\cal J}_2^{(0)} = \frac{q_2}{4 \pi} \frac{1}{2}.
    \label{eq:rotxelements}
\end{equation}
Eq (\ref{eq:phantomj2computed}) and (\ref{eq:rotxelements}) yield
\begin{equation}
    {\cal J}_2 = - \frac{q_2}{4 \pi}
    \left[
    \begin{array}{c}
    \sqrt{3/8} \\
    0 \\
    1/2 \\
    0 \\
    \sqrt{3/8}    
    \end{array} \right] = - \frac{q_2}{4 \pi} v^{(y)}_{2,0}.
    \label{eq:phantomj2explicit}
\end{equation}
The last equality above shows that ${\cal J}_2$ is proportional to
the eigenvector of $L^{(2)}_y$ that is invariant under rotations about
the $y$-axis. 

We now compute the ingredients that make up the matrix element
${\cal M}_2$ given by eq (\ref{eq:matrixelement}). 
From eq (\ref{eq:phantomj2explicit}) it follows that 
\begin{equation}
    {\cal J}_2^\dagger v^{(y)}_{2 \mu} = - \frac{q_2}{4 \pi} 
    \label{eq:calj2daggervy}
\end{equation}
for $\mu = 0$ and zero for other values of $\mu$. Using eq (\ref{eq:innerquadrupolemoments})
for $J^{(0)}_2$ and the eigenvectors $v^{(y)}_{2 \mu}$ tabulated below
we find
\begin{eqnarray}
v^{(y) \dagger}_{2, \pm 2} J_2^{(0)} & = & - \frac{5 \sqrt{6}}{8} e_K^2, \nonumber \\
v^{(y) \dagger}_{2, 0} J_2^{(0)} & = & \frac{1}{8} (4 + 6 e_K^2 ).
\label{eq:vydaggerj2}
\end{eqnarray}
Making use of the tabulated eigenvectors and eq (\ref{eq:inclinationfactor}) 
we can also compute the relevant components of the inclination factor 
\begin{eqnarray}
    {\cal I}_{0, \pm 2} & = & \frac{1}{4} \sqrt{ \frac{3}{2} } 
    \left[ \cos (2 i_K ) - 1 \right], 
    \nonumber \\
    {\cal I}_{0, 0} & = & \frac{1}{4} + \frac{3}{4} \cos (2 i_K).
    \label{eq:quadrupoleinclination}
\end{eqnarray}

Combining the results in eqs (\ref{eq:calj2daggervy}), (\ref{eq:vydaggerj2})
and (\ref{eq:quadrupoleinclination}) in accordance with eqs (\ref{eq:mondisturb}) and
(\ref{eq:matrixelement}) we finally obtain the $\ell = 2$ or quadrupole term in the
disturbing function
\begin{equation}
    {\cal R}_Q = \frac{G m_K M_\odot}{R_M} \left( \frac{a_K}{R_M} \right)^2 
    \frac{q_2}{32 \pi} {\cal S}_Q
    \label{eq:quadrupoledisturbmond}
\end{equation}
with
\begin{eqnarray}
    {\cal S}_Q & = & -2 - 3 e_K^2 + 15 e_K^2 \cos (2 \omega_K) 
    + 6 \cos^2 i_K 
\nonumber \\
    & & + 9 e_K^2 \cos^2 i_K - 15 e_K^2 \cos(2 \omega_K) \cos^2 i_K.
    \nonumber \\
    \label{eq:scaledquadrupoledisturbmond}
\end{eqnarray}
Eqs (\ref{eq:quadrupoledisturbmond}) and (\ref{eq:scaledquadrupoledisturbmond}) are 
the key formulae in this section.

We now briefly consider the disturbing function that describes the effect of planet
nine on the dynamics of a KBO in the quadrupole approximation. We assume that
the orbit of planet nine lies in the $z$-$x$ plane with the sun at the focus and
the perihelion on the positive $z$-axis. We use the $z$-$x$ plane as the reference
plane and assume that the orbit of planet nine remains fixed for the time duration
of interest. The reference configuration of the KBO orbit is one wherein it lies in the
reference plane with the perihelion along the positive $z$-axis. The configuration
$(\omega_K, i_K, \Omega_K)$ corresponds to giving the orbit a $y$-$z$-$y$ sequence
of rotations by the angles $\omega_K$, $i_K$ and $\Omega_K$ respectively. 
Hence the quadrupolar disturbing function is given by the $\ell = 2$ term in
eq (\ref{eq:mondisturb}) with the replacements $M_\odot \rightarrow m_9$ and
$R_M \rightarrow a_9$. ${\cal M}_2$ is given by eq (\ref{eq:prematrixelement}) 
with $J_2^{(0)}$ still given by eq (\ref{eq:innerquadrupolemoments}) but with
${\cal J}_2$ given by eq (\ref{eq:planetnineouterquadrupole}) not by 
eq (\ref{eq:phantomj2explicit}). Comparing eq (\ref{eq:planetnineouterquadrupole}) 
to eq (\ref{eq:phantomj2explicit}) we see that the only change in ${\cal M}_2$ is
the replacement 
\begin{equation}
    \frac{q_2}{4 \pi} \rightarrow - \frac{1}{2} \frac{1}{(1 - e_9^2)^{3/2}}.
    \label{eq:mondtonine}
\end{equation}
Hence the disturbing function for planet nine acting on the KBO to quadrupole
order is given by making the same replacement in eq (\ref{eq:quadrupoledisturbmond})
leading to 
\begin{equation}
    {\cal R}_Q^{(9)} = - \frac{G m_K m_9}{a_9} \left( \frac{a_K}{a_9} \right)^2 
    \frac{1}{16} \frac{1}{(1 - e_9^2)^{3/2}} {\cal S}_Q.
    \label{eq:quadrupoledisturbnine}
\end{equation}
Here ${\cal S}_Q$ is still given by eq (\ref{eq:scaledquadrupoledisturbmond}).

On long time scales it is not valid to treat the orbit of planet nine as fixed.
It should evolve slowly in response to perturbations by the four known giant planets. 
In the analysis of the interaction of planet nine and a KBO it is therefore 
better to use a fixed plane such as the mean ecliptic as the reference plane
rather than the orbital plane of planet nine as we have done above. 
The disturbing function would then be a function of the orbital elements
of both the KBO and planet nine. We have derived such expressions for 
the disturbing function to quadrupole and octupole order, and in principle the
expressions can be extended to arbitrary orders using our computational
scheme. We will present this analysis elsewhere 
since we do not 
need these expressions for our present purpose. Similar expressions have been
obtained by \cite{mardling} but are not available in the literature. 

Another useful variation of eq (\ref{eq:quadrupoledisturbnine}) 
to consider is the disturbing function that describes
the effect of Neptune or one of the other giant planets on the motion of a KBO
in the secular approximation. The same expression 
applies as for the interaction of a KBO with planet nine but with the following 
replacements in eqs (\ref{eq:quadrupoledisturbnine}) and (\ref{eq:scaledquadrupoledisturbmond}): 
$ (m_K, a_K, e_K) \rightarrow (m_8, a_8, e_8)$ and
$(m_9, a_9, e_9) \rightarrow (m_K, a_K, e_K)$ since Neptune replaces the
KBO as the object in the inner orbit and the KBO replaces planet nine 
as the outer object. Bearing in mind that, to a good approximation,
the giant planets have essentially circular orbits in a common plane,
their combined effect on a KBO to quadrupole order is described by the
following well-known approximate disturbing function
\begin{equation}
    {\cal R}_Q^{({\rm G})} = \frac{G m_K}{a_K^3} \frac{1}{8} 
    \frac{(1 - 3 \cos^2 i_K)}{(1 - e_K^2)^{3/2}}
        \sum_{i=5,6,7,8} m_i a_i^2.
    \label{eq:giantdisturb}
\end{equation}

Now let us return to the disturbing function that describes the effect
of the galactic field on a KBO in MOND and work out the $\ell = 3$
octupole term in eq (\ref{eq:mondisturb}). The derivation is similar
to that of the quadrupole and for the sake of brevity we only provide
the final result.
\begin{equation}
    {\cal R}_{{\rm oct}} = - \frac{G m_K M_\odot}{R_M} 
    \left( \frac{a_K}{R_M} \right)^3 \frac{q_3}{16 \pi} {\cal S}_{{\rm oct}}.
    \label{eq:octupolemondisturb}
\end{equation}
Here $q_3$ is given by eq (\ref{eq:phantomultipole}) and
\begin{eqnarray}
    {\cal S}_{{\rm oct}} & = & \frac{175}{16} e_K^3 \sin^3 i_K \sin 3 \omega_K 
    \nonumber \\
    & & + \frac{15}{16} ( 4 e_K + 3 e_K^3 ) \sin \omega_K 
    (4 \sin i_K - 5 \sin^3 i_K).
    \nonumber \\
    \label{eq:scaledoctupole}
\end{eqnarray}

To conclude this section we tabulate the eigenvectors of $L_x$
and $L_y$ that are used in the calculations above. 
The eigenvectors of $L_y^{(2)}$ are denoted 
$v^{(y)}_{2 \mu}$ for $ - 2 \leq \mu \leq +2$. The first column of the
following matrix is $v^{(y)}_{2, 2}$, the second column is $v^{(y)}_{2,1}$ and so on
to the last column which corresponds to $v^{(y)}_{2, -2}$.
These eigenvectors are orthonormal and satisfy 
$v^{{(y) \dagger}}_{2, \mu} v^{(y)}_{2, \nu} = \delta_{\mu, \nu}$ and
hence the eigenmatrix below is unitary. Similar remarks apply to all
the eigenmatrices below. 
\begin{equation}
    \left[ \begin{array}{ccccc}
    1/4 & 1/2 & \sqrt{3/8} & 1/2 & 1/4 \\
    i/2 & i/2 & 0 & - i/2 & -i/2 \\
    - \sqrt{6}/4 & 0 & 1/2 & 0 & - \sqrt{6}/4 \\
    - i/2 & i/2 & 0 & -i/2 & i/2 \\
    1/4 & - 1/2 & \sqrt{3/8} & - 1/2 & 1/4 \\
    \end{array} \right]
    \label{eq:l5y}
\end{equation}

The eigenvectors of $L_x^{(2)}$ are 
\begin{equation}
    \left[ \begin{array}{ccccc} 
    1/4 & -i/2 & - \sqrt{3/8} & i/2 & 1/4 \\
    1/2 & -i/2 & 0 & -i/2 & -1/2 \\
    \sqrt{6}/4 & 0 & 1/2 & 0 & \sqrt{6}/4 \\
    1/2 & i/2 & 0 & i/2 & -1/2 \\
    1/4 & i/2 & - \sqrt{3/8} & -i/2 & 1/4 \\
    \end{array} \right]
    \label{eq:l5x}
\end{equation}

For calculation of the octupole we need the eigenvectors of $L_y^{(3)}$
tabulated below. 
\begin{equation}
    \frac{1}{8} \left[ \begin{array}{ccccccc} 
    i & \sqrt{6} & - i \sqrt{15} & - \sqrt{20} & i \sqrt{15} & \sqrt{6} & - i \\
    - \sqrt{6} & 4 i & \sqrt{10} & 0 & \sqrt{10} & - 4 i & - \sqrt{6} \\
    - i \sqrt{15} & - \sqrt{10} & - i & - \sqrt{12} & i & - \sqrt{10} & i \sqrt{15} \\
    \sqrt{20} & 0 & \sqrt{12} & 0 & \sqrt{12} & 0 & \sqrt{20} \\
    i \sqrt{15} & - \sqrt{10} & i & - \sqrt{12} & - i & - \sqrt{10} & - i \sqrt{15} \\
    - \sqrt{6} & - 4 i & \sqrt{10} & 0 & \sqrt{10} & 4 i & - \sqrt{6} \\
    - i & \sqrt{6} & i \sqrt{15} & - \sqrt{20} & - i \sqrt{15} & \sqrt{6} &  i \\
    \end{array}
    \right]
    \label{eq:l7y}
\end{equation}

Finally the eigenvectors of $L_x^{(3)}$ are the following. 
\begin{equation}
    \frac{1}{8} \left[ \begin{array}{ccccccc}
    1 & - i \sqrt{6} & - \sqrt{15} & i \sqrt{20} & \sqrt{15} & - i \sqrt{6} & - 1 \\
    \sqrt{6} & - 4 i & - \sqrt{10} & 0 & - \sqrt{10} & 4 i & \sqrt{6} \\
    \sqrt{15} & - i \sqrt{10} & 1 & - i \sqrt{12} & -1 & - i \sqrt{10} & - \sqrt{15}  \\
    \sqrt{20} & 0 & \sqrt{12} & 0 & \sqrt{12} & 0 & \sqrt{20} \\
    \sqrt{15} & i \sqrt{10} & 1 & i \sqrt{12} & - 1 & i \sqrt{10} & - \sqrt{15} \\
    \sqrt{6} & 4 i & - \sqrt{10} & 0 & - \sqrt{10} & - 4 i & \sqrt{6} \\
    1 & i \sqrt{6} & - \sqrt{15} & - i \sqrt{20} & \sqrt{15} & i \sqrt{6} & - 1 \\
    \end{array} \right]
    \label{eq:l7x}
\end{equation}

\section{Dynamics} 

We now analyze the effect of MOND on the dynamics of a KBO. 
In Lagrangian or Hamiltonian mechanics the state of the KBO would
be specified by its position and velocity or by its position and canonical 
momentum---a total of six dynamical variables or three degrees of freedom. 
In celestial mechanics it is more convenient to work with a noncanonical set
of variables, the orbital elements $(a_K, e_K, \omega_K, i_K, \Omega_K, f_K)$.
$a_K$ and $e_K$ specify the shape of the instantaneous orbit (the orbit that
the KBO would pursue if all influences except the Newtonian gravity of the
sun were turned off); $(\omega_K, i_K, \Omega_K)$ the orientation of the orbit; 
and $f_K$ the location of the KBO along the orbit. To be precise $f_K$ is
the angle between the position vector of the KBO and the apsidal vector 
joining the sun to the perihelion. The dynamics of the orbital elements
is determined by the disturbing function via Lagrange's equations. 

In the secular approximation we focus on the behavior of the orbital elements
on time scales that are long compared to the orbital period of the instantaneous
orbit. In this approximation the disturbing function is time averaged over the orbit 
and the anomaly $f_K$ is eliminated from the equations. Furthermore it turns out that 
in the secular approximation the semimajor axis $a_K$ is conserved and may be 
regarded as a fixed parameter \citep{elliotplusmcdermott}. 
Hence in the secular approximation there are only four
dynamical variables $(e_K, \omega_K, i_K, \Omega_K)$ and
two degrees of freedom. The dynamics of these variables are 
governed by the following Lagrange equations.
\begin{eqnarray}
m_K \frac{d e_K}{d t} & = & - \frac{1}{n_K a_K^2} \sqrt{1 - e_K^2} 
    \frac{\partial {\cal R}}{\partial \omega_K} \nonumber \\
m_K \frac{d \Omega_K}{d t} & = & \frac{1}{n_K a_K^2} \frac{1}{\sqrt{1 - e_K^2}} 
    \frac{1}{\sin i_K} \frac{ \partial {\cal R}}{\partial i_K} \nonumber \\
m_K \frac{d \omega_K}{dt} & = & \frac{1}{n_K a_K^2} \frac{\sqrt{1 - e_K^2}}{e_K} 
    \frac{\partial {\cal R}}{\partial e_K} - 
    \frac{1}{n_K a_K^2} \frac{\cot i_K}{\sqrt{1 - e_K^2}} 
    \frac{\partial {\cal R}}{\partial i_K} \nonumber \\
m_K \frac{d i_K}{d t} & = & \frac{1}{n_K a_K^2} \frac{1}{\sqrt{1 - e_K^2}}
    \left( \cot i_K \frac{\partial R}{\partial \omega_k} 
    - \frac{1}{\sin i_K} \frac{\partial {\cal R}}{\partial \Omega_K} 
    \right). 
    \nonumber \\
    \label{eq:lagrangeqs}
\end{eqnarray}
Here $n_K = ( G M_\odot / a_K^3 )^{1/2}$ is the mean angular frequency of a Kepler
orbit of semimajor axis $a_K$. 

This is a formidable set of coupled nonlinear differential equations, but for our
problem a great simplification occurs due to symmetry, allowing an exact solution. 
First, it is easy to verify that the disturbing function ${\cal R}$ itself is 
a conserved quantity, by observing that
\begin{equation}
    \frac{d {\cal R}}{d t} =
    \frac{\partial {\cal R}}{\partial e_K} \frac{d e_K}{dt} +
    \frac{\partial {\cal R}}{\partial \Omega_K} \frac{d \Omega_K}{d t} + 
    \frac{\partial {\cal R}}{\partial \omega_K} \frac{d \omega_K}{d t} +
    \frac{\partial {\cal R}}{\partial i_K} \frac{d i_K}{d t}
    \label{eq:drdt},
\end{equation}
and making use of Lagrange's equations (\ref{eq:lagrangeqs}). 
Second, due to cylindrical symmetry about the axis joining the sun 
to the center of the galaxy, ${\cal R}$ is independent of $\Omega_K$.
More crucially the symmetry implies that the axial component of the
angular momentum is conserved. For a Keplerian orbit the 
magnitude of the angular momentum is given by
$[ G M_\odot m_K^2 a_K (1 - e_K^2) ]^{1/2}$ in terms of the
orbital elements. The component along the symmetry axis is 
$[ G M_\odot m_K^2 a_K (1 - e_K^2) ]^{1/2} \cos i_K$. Discarding
unimportant constants it follows that the quantity
\begin{equation}
    h = \sqrt{ 1 - e_K^2 } \cos i_K
    \label{eq:hdefined}
\end{equation}
is conserved. It is easy to verify explicitly by use of 
Lagrange's eqs (\ref{eq:lagrangeqs}) that $h$ is conserved if
${\cal R}$ is independent of $\Omega_K$. We will call $h$ the
dimensionless axial angular momentum in the following discussion. 

Hence our problem is integrable. There are four dynamical variables,
$(e_K, \omega_K, i_K, \Omega_K)$ and hence two degrees of freedom. 
There are also two conserved quantities: ${\cal R}$ and $h$. 

It is useful to make the following observations based on the
definition of $h$. (i) Since $0 \leq e_K \leq 1$ and $0 \leq i_K \leq \pi$
it follows that $|h| \leq 1$ with negative values of $h$ corresponding to
$\pi/2 < i_K \leq \pi$. (ii) For a given $h$, the eccentricity
lies in the range $0 \leq e_K \leq \sqrt{1 - h^2}$. (iii) For $h > 0$,
the inclination lies in the range $0 \leq i_K \leq i_{{\rm max}}$;
for $h < 0$, in the range $\pi - i_{{\rm max}} \leq i_K \leq \pi$.
Here $i_{{\rm max}} = \cos^{-1} |h|$ and it lies in the range
$0 \leq i_{{\rm max}} \leq \pi/2$. (iv) Obviously if $h=0$ then
either $e_K = 1$ or $i_K = \pi/2$. (v) If $h = \pm 1$ then $e_K = 0$
and $i_K = 0$ or $\pi$ respectively.

To gain some qualitative insight into the KBO dynamics we now work 
in the quadrupole approximation to the disturbing function given in 
eqs (\ref{eq:quadrupoledisturbmond}) and (\ref{eq:scaledquadrupoledisturbmond}).
We use eq (\ref{eq:hdefined}) to eliminate $i_K$ to obtain ${\cal S}_Q$
as a function of $(e_K, \omega_K)$. The result is
\begin{eqnarray}
    {\cal S}^{{\rm eff}}_Q & = & - 2 - 9 h^2 + 15 h^2 \cos 2 \omega_K \nonumber \\
    & & + (15 \cos 2 \omega_K - 3) e_K^2 + 15 h^2 (1 - \cos 2 \omega_K) \frac{1}{1 - e_K^2}.
    \nonumber \\
    \label{eq:landscape}
\end{eqnarray}
Here $0 \leq e_K \leq \sqrt{1-h^2}$ and $0 \leq \omega_K < 2 \pi$. 
The conservation of ${\cal R}_Q$ implies that the KBO trajectories 
will lie along contours of fixed ${\cal S}^{{\rm eff}}_Q$ in the $(e_K, \omega_K)$ plane.
Hence we can visualize the dynamics by plotting the contours of fixed 
${\cal S}^{{\rm eff}}_Q$.

It is easy to see that for fixed $e_K$, the scaled disturbing function 
${\cal S}^{{\rm eff}}_Q$ oscillates as a function of $\omega_K$ with minima at $\omega_K = \pi/2$
and $3 \pi/2$ and maxima at $0$ and $\pi$. This is facilitated by rewriting 
${\cal S}^{{\rm eff}}_Q$ in the form
\begin{eqnarray}
    {\cal S}^{{\rm eff}}_Q & = & - 2 - 9 h^2 - 3 e_K^2 + \frac{15 h^2}{1 - e_K^2} \nonumber \\
    & & + 15 \left( h^2 + e_K^2 - \frac{h^2}{1-e_K^2} 
    \right) \cos 2 \omega_K
    \label{eq:seascape}
\end{eqnarray}
and demonstrating that the coefficient of $\cos 2 \omega_K$ is positive for the
entire range $0 < e_K^2 < 1 - h^2$. It is also easy to see that 
${\cal S}^{{\rm eff}}_Q = -2 + 6 h^2$ for $e_K = 0$ and ${\cal S}^{{\rm eff}}_Q = 10 - 6 h^2$
for $e_K^2 = 1 - h^2$. Remarkably, we see that ${\cal S}^{{\rm eff}}_Q$ is independent
of $\omega_K$ for the extremal values of $e_K$. We also observe that 
\begin{equation}
    {\cal S}^{{\rm ef}}_Q (e^2_K \rightarrow 1 - h^2) - {\cal S}^{{\rm eff}}_Q (e_K^2 \rightarrow 0) 
    = 12 (1 - h^2) \geq 0.
    \label{eq:extremalvals}
\end{equation}

Next let us consider the behavior of ${\cal S}^{{\rm eff}}_Q$ as a function of $e_K^2$ for
fixed $\omega_K$. For this purpose it is better to revert to the form
in eq (\ref{eq:landscape}). We observe that the last term in ${\cal S}^{{\rm eff}}_Q$ with the factor
$1/(1 - e_K^2)$ is monotonically increasing with $e_K^2$ since its coefficient
is positive. For the $e_K^2$ term we can distinguish two cases. For 
$\cos 2 \omega_K > 1/5$ it is monotonically increasing with $e_K^2$ since
its coefficient is positive. For $\cos 2 \omega_K \leq 1/5$ it is not monotonically
increasing. Hence we can definitely state that ${\cal S}^{{\rm eff}}_Q$ is monotonically increasing
as a function of $e_K^2$ for all $\omega_K$ such that $\cos 2 \omega_K > 1/5$. 
But for $\cos \omega_K \leq 1/5$ further analysis is needed. These values of
$\omega_K$ correspond to two equal intervals centered on $\omega_K = \pi/2$
and $\omega_K = 3 \pi/2$ with a width greater than $\pi/2$ but less than $\pi$. 
In these intervals we will see that there are two possible behaviors as a function
of $e_K^2$. Either ${\cal S}^{{\rm eff}}_Q$ can remain monotonically increasing or it can first
decrease to a minimum and then increase again as $e_K^2$ varies from
$0$ to $1 - h^2$. To determine the actual behavior that is obtained we 
compute $\partial {\cal S}^{{\rm eff}}_Q / \partial e_K^2$ and find it is equal to
zero for $e_K^2 \rightarrow \overline{e}^2$ where
\begin{equation}
    \overline{e}^2 = 1 - | h | \sqrt{ \frac{ 1 - \cos 2 \omega_K }{(1/5) - \cos 2 \omega_K} }.
    \label{eq:minimum}
\end{equation}
Obviously we need $\cos 2 \omega_K < 1/5$ for $\overline{e}^2$ to be real but in fact
we need to impose the more stringent conditions that $0 \leq \overline{e}^2 \leq 1 - h^2$. 
The condition that $\overline{e}^2 \leq 1 - h^2$ 
is automatically fulfilled for $(1/5) - \cos 2 \omega_K > 0$
as we have assumed. But the condition that $\overline{e}^2 \geq 0$ translates to
\begin{equation}
    \cos 2 \omega_K \leq 1 - \frac{4}{5} \frac{1}{1 - h^2}.
    \label{eq:nontrivial}
\end{equation}
It is easy to see that the right hand side of eq (\ref{eq:nontrivial}) decreases monotonically
from $1/5$ to $-1$ as $h^2$ goes from $0$ to $3/5$ and is less than $-1$ for $h^2 > 3/5$.

In summary, we find the following behavior for ${\cal S}^{{\rm eff}}_Q$ considered as a function of
$e_K^2$ for fixed $\omega_K$. (i) For $3/5 < h^2 \leq 1$, we find ${\cal S}^{{\rm eff}}_Q$ is monotonic 
increasing for all $\omega_K$. (ii) For $0 \leq h^2 \leq 3/5$ we find that provided
eq (\ref{eq:nontrivial}) is satisfied, 
${\cal S}^{{\rm eff}}_Q$ shows non-monotonic behavior: it decreases for $0 \leq e_K^2 < \overline{e}^2$
and increases for $\overline{e}^2 < e_K^2 \leq 1 - h^2$ with a minimum at $e_K^2 = \overline{e}^2$
where $\overline{e}^2$ is given by eq (\ref{eq:minimum}). The range of $\omega_K$ defined by
eq (\ref{eq:nontrivial}) corresponds to two intervals centered about $\omega_K = \pi/2$ 
and $\omega = 3 \pi/2$. Outside of this range of $\omega_K$ values, ${\cal S}_Q$ remains
monotonic even for $0 \leq h^2 \leq 3/5$. Crucially then we find that ${\cal S}^{{\rm eff}}_Q$ has two 
global minima at $(e_K, \omega_K)$ given by $(e_C, \pi/2)$ and $(e_C, 3 \pi/ 2)$ where
\begin{equation}
    e_C^2 = 1 - |h| \sqrt{ \frac{5}{3} }.
    \label{eq:ec}
\end{equation}
Eq (\ref{eq:ec}) is obtained from eq (\ref{eq:minimum}) by setting $\omega_K \rightarrow 
\pi/2$ or $3 \pi/2$. This result for the location of the minimum is one of the key
results of this section.

We are now ready to plot contours of fixed
${\cal S}^{{\rm eff}}_Q$. It is convenient to define $\Delta {\cal S}^{{\rm eff}}_Q = {\cal S}^{{\rm eff}}_Q - {\cal S}^{{\rm eff}}_Q 
( e_K^2 \rightarrow 0)$. Making use of eq (\ref{eq:scaledquadrupoledisturbmond}) we find
explicitly that
\begin{eqnarray}
\Delta {\cal S}^{{\rm eff}}_Q & = & - 15 (1 - \cos 2 \omega_K ) + (15 \cos 2 \omega_K - 3) e_K^2
    \nonumber \\
    & & 
    + \frac{15 h^2}{(1 - e_K^2)} (1 - \cos 2 \omega_K ) 
        \label{eq:deltasq}
\end{eqnarray}
Evidently $\Delta {\cal S}^{{\rm eff}}_Q = 0$ for $e_K^2 = 0$ and $\Delta {\cal S}^{{\rm eff}}_Q = 12 (1 - h^2)$ for 
$e_K^2 = 1 - h^2$. (i) For $h^2 > 3/5$, $\Delta {\cal S}^{{\rm eff}}_Q$ increases monotonically 
with $e_K^2$ for fixed $\omega_K$ and oscillates with $\omega_K$ for fixed
$e_K$ as discussed above. Hence we expect the contours of fixed $\Delta {\cal S}_Q^{{\rm eff}}$ 
to be wavy lines with maxima at $\omega_K = \pi/2$ and $3 \pi / 2$ and minima 
at $\omega_K = 0, \pi$ and $2 \pi$. The contours for different values of $\Delta {\cal S}^{{\rm eff}}_Q$
between zero and the maximal value of $12 ( 1 - h^2 )$ may be plotted exactly by 
solving eq (\ref{eq:deltasq}) for $e_K^2$ as a function of $\omega_K$. This is a
quadratic equation and hence we have to choose the root by either verifying the pattern
of maxima and minima noted above or by continuity with the expected result as 
$e_K^2 \rightarrow 0$. (ii) For $0 \leq h^2 \leq 3/5$ it is convenient to first plot the
contour corresponding to $\Delta {\cal S}^{{\rm eff}}_Q = 0$. Eq (\ref{eq:deltasq}) has the trivial
solution $e_K^2 = 0$ for $\Delta {\cal S}^{{\rm eff}}_Q = 0$ and a nontrivial solution 
\begin{equation}
    e_K^2 = 1 - h^2 \frac{1 - \cos 2 \omega_K}{1/5 - \cos 2 \omega_K}.
    \label{eq:dome}
\end{equation}
We require that $0 \leq e_K^2 \leq 1 - h^2$. Imposing these
conditions on the right hand side of eq (\ref{eq:dome}) we find as expected that the
nontrivial solution exists over the range of $\omega_K$ that satisfy eq (\ref{eq:nontrivial}). 
This range corresponds to two equal intervals, one centered about $\omega_K = \pi/2$ 
and the other centered about $\omega_K = 3 \pi / 2$. Plotting the trivial and nontrivial
solutions in red we see from 
Fig 2(b) in the main body of the paper that they define two 
dome shaped regions. $\Delta {\cal S}^{{\rm eff}}_Q > 0$ outside the dome shaped regions 
and $\Delta {\cal S}^{{\rm eff}}_Q < 0$ inside the domes. In particular the domes contain
the minima of $\Delta {\cal S}^{{\rm eff}}_Q$ which are located at $(e_K, \omega_K) 
\rightarrow (e_C, \pi/2$ or $3 \pi / 2)$ where $e_C$ is given by eq (\ref{eq:ec}).
At the minima it is easy to compute using eq (\ref{eq:deltasq}) that
\begin{equation}
    \Delta {\cal S}^{{\rm eff}}_Q = -30 \left( \sqrt{ \frac{3}{5} } - |h| \right)^2.
    \label{eq:minsq}
\end{equation}
Thus the minima are deepest for $|h| \rightarrow 0$. We expect the contours to
be wavy lines outside the domes and loops within. 
To actually plot the
contours we proceed as in case (i) for the region outside the domes, 
plotting the appropriate solution to eq (\ref{eq:deltasq}). But inside
the dome we plot both solutions to eq (\ref{eq:deltasq}) over the range of
$\omega_K$ for which these solutions are real. Those plots join smoothly
to form the expected loops. The results are shown in fig 2 
in the main body of the paper.

Dynamically the phase space flow therefore corresponds to
precession along the wavy lines for $h^2 > 3/5$. For $h^2 < 3/5$
the phase space breaks up into two regions: one in which the trajectories correspond
to precession (i.e. monotonic increase of $\omega_K$) and the other 
in which the trajectories encircle the minima which are fixed points of the dynamics. 

For the record we provide the specific parameters used to generate Fig. 2 in the main body
of the paper. 
In (a) we chose $h = 0.9 > \sqrt{3/5}$. The allowed range of $\Delta {\cal S}^{{\rm eff}}_Q$ 
is $0 \leq \Delta {\cal S}^{{\rm eff}}_Q \leq 12 (1 - h^2) = 2.28$. The plotted contours correspond to
$\Delta {\cal S}^{{\rm eff}}_Q = 0.4, 0.8, 1.2, 1.6$ and $2.0$. In (b) we chose $h = 0.5 < \sqrt{3/5}$. 
Outside the domes $0 < \Delta {\cal S}^{{\rm eff}}_Q \leq 12 (1 - h^2) = 9$. The plotted contours
correspond to $\Delta {\cal S}^{{\rm eff}}_Q = 2, 4, 6$ and $8$. Inside the domes $-2.28\ldots 
\leq \Delta {\cal S}^{{\rm eff}}_Q < 0$. The lower limit comes from eq (\ref{eq:minsq}).
The plotted contours correspond to $\Delta {\cal S}^{{\rm eff}}_Q = -0.75, -1.5$ and $-2.25$.

Plotting the contours of ${\cal S}^{{\rm eff}}_Q$ determines the curves traced by the KBO
though the $(e_K, \omega_K)$ phase space but not the direction of the flow.
To obtain the direction we must return to Lagrange's equations (\ref{eq:lagrangeqs}).
Making use of the quadrupole approximation to ${\cal R}_Q$ in eq (\ref{eq:quadrupoledisturbmond}) 
and using the conservation of $h$ defined by eq (\ref{eq:hdefined}) we find
\begin{eqnarray} 
\frac{d e_K^2}{d t} & = & \frac{60}{T_K} e_K^2 \sqrt{1 - e_K^2} \sin 2 \omega_K 
    \left[ 1 - \frac{h^2}{1 - e_K^2} \right], \nonumber \\
\frac{d \omega_K}{d t} & = & \frac{1}{T_K} \frac{1}{\sqrt{1 - e_K^2}} \left[ 
- 6 (1 - e_K^2) + 30 (1 - e_K^2) \cos 2 \omega_K \right]
\nonumber \\
& & 
+ \frac{1}{T_K} \frac{1}{\sqrt{1 - e_K^2}}
\left[ 30 (1 - \cos 2 \omega_K ) 
\frac{h^2}{1 - e_K^2}
\right]. 
\nonumber \\
\label{eq:lagrangianphase}
\end{eqnarray}
Here $T_K$ is defined via
\begin{equation}
    \frac{1}{T_K} = n \left( \frac{a_K}{R_M} \right)^3 \frac{q_2}{32 \pi}
    \label{eq:tkdef}
\end{equation}
and is the natural time scale for MOND induced evolution of the KBO orbit. 
The first of these equations is sufficient to determine the direction of the
flow. We see that $d e_K^2 / d t \geq 0$ for $0 \leq \omega_K \leq \pi/2$
and $\pi \leq \omega_K \leq 3 \pi / 2$; and $d e_K^2 / d t < 0$ for 
$\pi/2 < \omega_K < \pi$ and $3 \pi / 2 < \omega_K < 2 \pi$. From this
we infer that the precession is from left to right in the figure 
($\omega_K$ increases monotonically with time) while the oscillations
are clockwise in the figure. 

It is easy to verify that if we set $(e_K, \omega_K)$ to the fixed point
values $(e_C, \pi/2$ or $3 \pi/2)$ on the right hand side of eq (\ref{eq:lagrangianphase})
then $d e_K^2 / dt = 0$ and $d \omega_K / d t = 0$ as expected. Linearizing
the equation of motion (\ref{eq:lagrangianphase}) about the fixed point yields
\begin{eqnarray}
\frac{d}{dt} \Delta e_K^2 & = & 
    \left( \frac{5}{3} \right)^{\frac{1}{4}}
    \frac{120 |h|^{1/2}}{T_K} 
    \left( \sqrt{\frac{3}{5}} - | h | \right) \left( \sqrt{\frac{5}{3}} - |h| \right) \Delta \omega_K
    \nonumber \\
\frac{d}{dt} \Delta \omega_K & = & - \left( \frac{3}{5} \right)^{\frac{1}{4}} 
    \frac{72}{T_K |h|^{1/2}}
\Delta e_K^2 
\nonumber \\
\label{eq:lin}
\end{eqnarray}
where $\Delta e_K^2 = e_C^2 - e_K^2$ and $ \Delta \omega_K = \omega_K - \pi/2$ or
$\Delta \omega_K = \omega_K - 3 \pi / 2$. 
From eq (\ref{eq:lin}) we see that the frequency of small oscillations about the fixed
point is 
\begin{equation}
    \frac{\sqrt{8640}}{T_K} \left( \sqrt{ \frac{3}{5} } - | h | \right)^{1/2} 
    \left( \sqrt{ \frac{5}{3} } - | h | \right)^{1/2}.
    \label{eq:smalloscfreq}
\end{equation}
We see that the frequency increases monotonically as $|h| \rightarrow 0$ reaching the 
maximum value of 
\begin{equation}
    \frac{24 \sqrt{15}}{T_K}.
    \label{eq:maxfreq}
\end{equation}

Thus far we have focussed on the dynamics of $e_K$ and $\omega_K$. The dynamics of 
$i_K$ are latched to those of $e_K$ by the conservation of $h$ via eq (\ref{eq:hdefined}). 
It is worth noting explicitly that at the fixed point $(e_C, \pi/2$ or $3 \pi/2)$
that 
\begin{equation}
    \cos i_K = \left( \frac{3}{5} \right)^{1/4} {\rm sgn} (h) \; | h |^{1/2}.
    \label{eq:cosifp}
\end{equation}
This is established using eq (\ref{eq:ec}) and $(\ref{eq:hdefined})$.
Here ${\rm sgn}(h) = +1$ for $h > 0$ and ${\rm sgn}(h) = -1$ for $h < 0$. 
It follows that $ e_C \rightarrow 1$ as $h \rightarrow 0$ and 
$i_K \rightarrow \pi/2$. More precisely
\begin{equation}
    i_K \approx \frac{\pi}{2} - \left( \frac{3}{5} \right)^{1/4} {\rm sgn} (h) \; |h|^{1/2}
    \label{eq:inclfp}
\end{equation}
for small $h$.

The dynamics of $\Omega_K$ is also fully determined by that of $e_K$ and $\omega_K$. 
To see this we return to the Lagrange equations (\ref{eq:lagrangeqs}) and make
use of the quadrupole approximations in eqs (\ref{eq:quadrupoledisturbmond})
and (\ref{eq:scaledquadrupoledisturbmond}) and of the conservation law
in eq (\ref{eq:hdefined}) to obtain
\begin{equation}
    \frac{d \Omega_K}{d t} = - \frac{1}{T_K} \frac{h}{(1 - e_K^2)} 
    \left[ 12 (1 - e_K^2) + 30 e_K^2 (1 - \cos 2 \omega_K) \right].
    \label{eq:Omegadyn} 
\end{equation}
We see that if $e_K$ and $\omega_K$ are known functions of time we can
readily integrate eq (\ref{eq:Omegadyn}) to obtain $\Omega_K$. It is 
also evident from eq (\ref{eq:Omegadyn}) that $ d \Omega_K / d t < 0$
for $h > 0$ and $d \Omega_K / d t > 0$ for $h < 0$. In either case 
$\Omega_K$ varies monotonically in time although it may not do so uniformly. 
To obtain the behavior of $\Omega_K$ when $(e_K, \omega_K)$ are at a fixed
point we set $e_K \rightarrow e_C$ and $\omega_K \rightarrow \pi/2$ or $3 \pi /2$
in eq (\ref{eq:Omegadyn}). We obtain
\begin{equation}
    \frac{d \Omega_K}{d t} = - \frac{1}{T_K} {\rm sgn}(h) \; \sqrt{ \frac{3}{5} } 
    \left[ 60 - 48 \sqrt{ \frac{5}{3} } | h | \right] 
    \label{eq:Omegadynfp}
\end{equation}
In other words we obtain uniform precession in $\Omega_K$ at the fixed point
with $d \Omega_K / d t = 0$ for the special case that $h = 0$. 

Thus far we have worked with the quadrupole approximation to the disturbing 
function. We now consider how the integrable dynamics above is modified
by the octupole correction given by eqs (\ref{eq:octupolemondisturb}) and
(\ref{eq:scaledoctupole}). In particular we are interested in how the two
fixed points located at $(e_C, \pi/2)$ and $(e_C, 3 \pi / 2)$ are affected. 
We find that the octupole term shifts the locations of these fixed points
slightly but more crucially it makes the fixed point at $(e_C, \pi/2)$ 
less stable than the one at $(e_C, 3 \pi / 2)$. Hence it is the latter fixed
point that will be dominant in shaping the anomalous structure of the Kuiper belt. 

We sketch briefly how these results are established. The octupole term respects
the cylindrical symmetry of the problem so $h$ remains conserved and we can
use eq (\ref{eq:hdefined}) to eliminate $i_K$ from eq (\ref{eq:scaledoctupole}).
Setting $(e_K, \omega_K)$ to the two fixed point values we see that ${\cal R}^{{\rm eff}}_{{\rm oct}}$
is positive at the first fixed point and negative at the second consistent with the idea
that the octupole term is destabilizing the first and stabilizing the second. 
In order to make this quantitative we expand ${\cal R}^{{\rm eff}}_{{\rm oct}}$ around each of the fixed
points to quadratic order. For the fixed point located at $(e_C, \pi/2)$ in the quadrupole
approximation, we find that
it shifts to a slightly higher value of $e_K$ and becomes less stable as evidenced by 
the frequency of small oscillations about the fixed point which becomes lower. For the fixed
point at $(e_C, 3 \pi / 2)$ in the quadrupole approximation, 
we find that it shifts to a slightly lower value of 
$e_K$ and becomes more stable as evidenced by the frequency of small oscillations 
which becomes higher. The formulae for the shifts in position and frequencies are long
and not very illuminating so for brevity we omit them. 

Now let us consider the effect of symmetry breaking perturbations on the integrable
MOND induced dynamics considered so far. The perturbations include non-secular terms
in the MOND disturbing function as well as secular and non-secular perturbations
caused by the giant planets. For
the long time dynamics the slow variation in the symmetry axis as the sun rotates around the galaxy 
must also be considered [for an analogous problem see \citep{banik2018b}]. 
According to Hamiltonian chaos theory \citep{percival} 
the phase space flow will become chaotic under these perturbations but 
the regular flow around the stable fixed point $(e_C, 3 \pi / 2)$ should 
persist especially for small $h$. Hence we predict that a population of 
KBOs should be found in orbits close to the stable fixed point for 
small $h$ (we will specify below the relevant values of $h$). This is
the central result of the paper. 

Let us describe more fully the key characteristics of the orbits of this
predicted KBO population that are stabilized by the galactic field in MOND.
Recall that we are using as the reference plane
not the mean ecliptic but rather the plane perpendicular to 
the direction to the center of the galaxy. Thus an orbit with
$\omega_K = 3 \pi /2$ and $i_K = \pi/2$ will have its apsidal
vector $\hat{{\boldsymbol \alpha}}_K$ pointing directly away from the center of the galaxy.
Here $\hat{{\boldsymbol \alpha}}_K$ is a unit vector that points from the sun to
the perihelion. 
This would be the case for a fixed point orbit
with $h = 0$. However as we will explain momentarily, we expect
$h$ to be small for reasons of stability, 
but greater than a minimum threshold value $h_{{\rm min}}$
that depends on the seminar major axis $a_K$. 
Thus we expect the orbits to have $i_K \approx \pi/2$ and
for the apsidal vector $\hat{{\boldsymbol \alpha}}_K$ to be inclined at a small angle 
$(\pi/2 - i_K)$ relative to $- \hat{{\bf n}}_G$.
Here $\hat{{\bf n}}_G$ is the unit vector that points from the sun 
to the center of the galaxy. 

Let us denote by $A_K$ the angle between $\hat{{\boldsymbol \alpha}}_K$ and 
$\hat{{\bf n}}_G$. We can estimate the expected range of $A_K$ as follows. 
For a given $h$ in the range $0 \leq h^2 \leq 3/5$, the orbital 
elements $(e_K, \omega_K)$ of the stable fixed point are 
$(e_C, 3 \pi / 2)$ in the quadrupole approximation. Here
$e_C$ is given by eq (\ref{eq:ec}). Making use of eq (\ref{eq:hdefined}) 
and (\ref{eq:ec})
we conclude that 
\begin{equation}
    \cos i_K = \left( \frac{3}{5} \right)^{1/4} {\rm sgn}\;(h) | h |^{1/2}
    \label{eq:fpincl}
\end{equation}
at the fixed point. Now $A_K = \pi/2 + i_K$ which leads to 
\begin{equation}
   \hat{{\boldsymbol \alpha}}_K \cdot \hat{{\bf n}}_G = \cos A_K = 
    - \sqrt{1 - \left( \frac{3}{5} \right)^{1/2} |h| }
    \label{eq:cosak}
\end{equation}
Assuming that $0 < |h| \leq (3/5)^{1/2}$ it follows that
$129^\circ < A_K \leq 180^\circ$. Taking into account that
the lower bound on $|h|$ is $h_{{\rm min}}$ we can refine
our estimate to read
\begin{equation}
    129^\circ < A_K < \cos^{-1} \left( - \sqrt{ 1 - \left( \frac{3}{5} \right)^{1/2} 
    h_{{\rm min}} }\right)
    \label{eq:akrange}
\end{equation}
We will see below that for $a_K \gg a_8$, $h_{{\rm min}} \rightarrow 0$
and the upper limit of the range of expected $A_K$ approaches $180^\circ$.
By contrast in the absence of MOND the galactic field has no significant
effect on KBOs and there should be no correlation between $\hat{{\boldsymbol \alpha}}_K$
and $\hat{{\bf n}}_G$. 

Next let us explain the origin of the bound $h_{{\rm min}}$. The closest approach 
of the KBO to the sun is $a_K (1 - e_K)$. For the orbit to be detached from the
inner solar system we need
\begin{equation}
    a_K (1 - e_K) > a_8.
    \label{eq:detach}
\end{equation}
Taking $e_K$ to be the fixed point value $e_C$ given by eq (\ref{eq:ec}) we find
that eq (\ref{eq:detach}) leads to the condition that $|h| > h_{{\rm min}}$ where
\begin{equation}
    h_{{\rm min}} = \sqrt{ \frac{3}{5} } \left[ 1 - \left( 1 - \frac{a_8}{a_K} \right)^2 \right].
    \label{eq:hmin}
\end{equation}
For $a_K \gg a_8$ we see that $h_{{\rm min}} \sim 2 \sqrt{3/5} a_8/a_K \rightarrow 0$
as $a_K \rightarrow \infty$ as noted above. 

Finally let us comment briefly on the orbit normal $\hat{{\mathbf l}}_K$. Obviously 
$\hat{{\mathbf l}}_K$ must be perpendicular to $\hat{{\boldsymbol \alpha}}_K$ 
and hence we expect $\hat{{\mathbf l}}_K$ to be roughly at $90^\circ$ to $\hat{{\mathbf n}}_G$
as well to the same extent that $\hat{{\boldsymbol \alpha}}_K$ is anti-aligned with
$\hat{{\mathbf n}}_G$. However due to precession of $\Omega_K$ we expect $\hat{{\mathbf l}}_K$
to be as random as possible consistent with the constraint that it is strictly perpendicular
to $\hat{{\boldsymbol \alpha}}_K$. Since $\hat{{\mathbf n}}_G$ lies almost in the ecliptic
plane it follows that the predicted population of KBOs may be found in the ecliptic plane
as well as at high inclinations to it. 

In summary the main result of this section is the prediction that there is a population of
KBOs with orbits clustered in the $(e_K, \omega_K)$ phase space near the stable fixed point
of the dynamics. The fixed point is at $(e_C, 3 \pi /2)$ in the quadrupole approximation
where $e_C$ is given by eq (\ref{eq:ec}). We have described the orbital characteristics
of this population and in the next section we will show that an observed population of
KBOs does indeed have orbits consistent with the predicted characteristics.


\section{Comparison to Data}

There are several well established families of Kuiper belt objects.
Many KBOs are in orbits that are locked in 2:1 or 3:2 resonances
with Neptune. The classical Kuiper belt objects have 
semi-major axes that lie between the two resonances noted above.
Centaurs penetrate the inner solar system but also recede far beyond 
the orbit of Neptune. The scattered disk consists of KBOs that recede 
far beyond the orbit of Neptune but have a perihelion distance comparable 
to that of Neptune. Recently a new subclass of this family has been 
recognized: objects that have perihelion distances well beyond the 
semimajor axis of Neptune and highly eccentric orbits that carry
them to the outer reaches of the solar system. Sedna \citep{sedna}
and 2012 VP$_{113}$ \citep{trujillo} were the first members of this subclass
to be discovered. We will therefore call this the Sedna family
of KBOs. 

The orbits of KBOs are for the most part compatible with dynamical 
models of the solar system based on eight planets and Newtonian gravity.
The Sedna family are a notable exception. The Planet Nine hypothesis was
introduced to explain the anomalous structure of their orbits.
Here we will explore the alternative hypothesis that this anomalous
structure is due to the galactic field in MOND. Because their
orbits are detached from the inner solar system the Sedna family
are a particularly clean and sensitive probe of MOND effects
and are the exclusive focus of this paper. However we note that the
Centaurs also pose a puzzle: they have a broad distribution 
of inclinations with many cases of very high inclination
and even retrograde motion. The dynamics of Centaurs are
chaotic and complicated by their penetration into the inner
solar system and therefore not considered further here. However
the finding in the previous section that MOND effects can lead
to orbits with very high inclinations is potentially significant
in this context and may be worth further consideration. 

In their comprehensive review of the planet nine hypothesis 
\cite {review} identify six members of the Sedna family that dynamical simulations reveal have stable orbits under the influence of the known planets and eight more that have metastable orbits. These fourteen objects were the trans Neptunian objects in the minor planet database of the International Astronomical Union whose orbital parameters satisfied the criteria that $a \geq 250$ au, $q \geq 30$ au (where $q$ denotes the perihelion distance) and $i \leq 40^\circ$ as of 10 October 2018. As of 21 June 2022 there are eight additional objects that meet these criteria in the IAU database. Table I lists the orbital parameters for all 22 KBOs: the fourteen discussed in the review by \cite{review} as well as the eight additional ones that are now known. These data are taken from the minor planet database; references to the discovery papers and the primary observational literature may be found on the website of the minor planet center. Note that for all 22 KBOs $q > a_8$ where $a_8 = 30.07$ au is the semimajor axis of Neptune’s nearly circular orbit. These KBOs are a good testing ground for the planet nine hypothesis and also for MOND effects. In the main body of the paper we limited ourselves to the six with stable orbits as they are likely the most clear cut members of the new population of KBOs that we predict based on MOND. Here also our main focus is on the same six KBOs but in addition we show that the orbits of the eight metastable objects and the eight new ones are also compatible with MOND.

 \begin{table} 
 \begin{ruledtabular}
 \begin{tabular}{lcccccc}
 Object & $\omega$ & $\Omega$ & $i$ & $e$ & $q$ (au) & $a$ (au) \\ \hline
 Sedna & 311.1$^\circ$ & 144.2$^\circ$ & 11.9$^\circ$ & 0.85 & 76.37 & 510.39 \\
 TG387$^a$ & 118.0$^\circ$ & 300.8$^\circ$ & 11.7$^\circ$ & 0.94 & 65.04 & 1031.49 \\
 2012 VP$_{113}$ & 293.5$^\circ$ & 90.7$^\circ$ & 24.1$^\circ$ & 0.69 & 80.39 & 258.27 \\
VN112$^b$ & 326.8$^\circ$ & 66.0$^\circ$ & 25.6$^\circ$ & 0.85 & 47.30 & 318.97 \\
GB174 & 347.0$^\circ$ & 130.9$^\circ$ & 21.6$^\circ$ & 0.86 & 48.61 & 336.67 \\
SR349 & 340.0$^\circ$ & 34.8$^\circ$ & 18.0$^\circ$ & 0.84 & 47.69 & 302.23 \\ \hline
RX245 & 64.6$^\circ$ & 8.6$^\circ$ & 12.1$^\circ$ & 0.90 & 45.73 & 448.49 \\
KG163 & 32.3$^\circ$ & 219.1$^\circ$ & 14$^\circ$ & 0.95 & 40.49 & 776.24 \\
GT50 & 129.3$^\circ$ & 46.1$^\circ$ & 8.8$^\circ$ & 0.88 & 38.48 & 324.66\\
TG422 & 285.6$^\circ$ & 112.9$^\circ$ & 18.6$^\circ$ & 0.92 & 35.55 & 468.98\\
FE72 & 133.5$^\circ$ & 337$^\circ$ & 20.7$^\circ$ & 0.98 & 36 & 1586.3\\
SY99 & 31.7$^\circ$ & 29.5$^\circ$ & 4.2$^\circ$ & 0.94 & 50.1 & 815.97\\
RF98 & 311.6$^\circ$ & 67.6$^\circ$ & 29.6$^\circ$ & 0.90 & 36.07 & 357.63\\
FT28 & 40.8$^\circ$ & 217.7$^\circ$ & 17.4$^\circ$ & 0.85 & 43.41 & 297.64\\ \hline
EU5 & 109.5$^\circ$ & 109.3$^\circ$ & 18.3$^\circ$ & 0.95 & 46.65 & 973.47\\
VM35 & 303.6$^\circ$ & 192.3$^\circ$ & 8.5$^\circ$ & 0.84 & 44.61 & 283.12\\
SD106 & 162.6$^\circ$ & 219.4$^\circ$ & 4.8$^\circ$ & 0.89 & 42.75 & 378.97\\
WB556 & 235.5$^\circ$ & 114.8$^\circ$ & 24.2$^\circ$ & 0.86 & 42.7 & 299.72\\
TU115 & 225.1$^\circ$ & 192.3$^\circ$ & 23.5$^\circ$ & 0.90 & 35.01 & 344.29\\
SL102 & 265.5$^\circ$ & 94.7$^\circ$ & 6.5$^\circ$ & 0.89 & 38.12 & 338.01\\
RA109 & 262.9$^\circ$ & 104.7$^\circ$ & 12.4$^\circ$ & 0.91 & 45.99 & 504.02\\
FL28 & 225$^\circ$ & 294.5$^\circ$ & 15.8$^\circ$ & 0.90 & 32.17 & 336.45\\

 \end{tabular}
 \end{ruledtabular}
 \caption{\label{kbodata} Orbital elements of KBOs of the Sedna family. The data are from the Minor Planet Database of the International Astronomical Union. As discussed in the review by \cite{review}, the first six have stable orbits under the influence of the known planets; the next eight are metastable. The final eight have been added to the database since the publication of the review (between 10 Oct 2018 and 21 Jun 2022). 
 $^a$ TG387 is named
 Leleakuhonua. $^b$ VN112 is named Alicanto. }
 \end{table}

The orientation of the orbits is specified relative to the
mean ecliptic plane which is taken to be the $X$-$Y$ plane. 
In the reference orientation the orbit of a KBO is assumed
to lie in the $X$-$Y$ plane with the perihelion along the positive
$X$ axis. The equation of the orbit in the reference configuration
is therefore
\begin{eqnarray}
X & = & a_K \cos \xi - a_K e_K \nonumber \\
Y & = & a_K \sqrt{ 1 - e_K^2 } \sin \xi \nonumber \\ 
Z & = & 0.
\label{eq:referenceorbit}
\end{eqnarray}
Here $0 \leq \xi < 2 \pi$ and $a_K$ and $e_K$ can be looked up from Table 1.
The actual orientation of the orbit is found by rotating it
about the $Z$ axis by $\omega$, then about the $X$ axis by $i$ 
and again the $Z$ axis by $\Omega$. After these transformations the equation of the
orbit is given by 
\begin{eqnarray}
X & = & (a_K \cos \xi - a_K e_K)( \cos \omega \cos \Omega - \sin \omega \sin \Omega \cos i)
\nonumber \\
& & + a_K \sqrt{1 - e_K^2} \sin \xi (- \sin \omega \cos \Omega - \cos \omega \sin \Omega \cos i)
\nonumber \\
Y & = & (a_K \cos \xi - a_K e_K)( \cos \omega \sin \Omega + \sin \omega \cos \Omega \cos i)
\nonumber \\
& & + a_K \sqrt{1 - e_K^2} \sin \xi ( - \sin \omega \sin \Omega + \cos \omega \cos \Omega \cos i)
\nonumber \\
Z & = & (a_K \cos \xi - a_K e_K)( \sin \omega \sin i) \nonumber \\
& & + a_K \sqrt{1 - e_K^2} \sin \xi (\cos \omega \sin i)
\label{eq:orbitdiagram}
\end{eqnarray}
In the ecliptic frame the latitude and longitude are denoted
$(\beta, \lambda)$. $\hat{{\mathbf n}}_G$ the unit vector that points in the direction
of the galactic center therefore has the components 
$(\cos \beta \cos \lambda, \cos \beta \sin \lambda, \sin \beta)$ 
with $\beta = -5.5^\circ $and $\lambda = 266.4^\circ $.


Using this information we can plot the orbits of all six KBOs
projected into the ecliptic plane and we can also project the
vector $\hat{{\mathbf n}}_G$ into the ecliptic plane. This leads to 
the plot shown in Fig. 3
in the main body of the paper. 
Note that the orbits as well as $\hat{{\mathbf n}}_G$ lie
almost in the ecliptic plane so this visualization gives
an accurate impression of their relative orientation. 
We see from this figure that the orbits
have a remarkable alignment of their apsidal vectors. This 
striking alignment was discovered and highlighted by 
\cite{trujillo} and \cite{batygin2016}. But we see now that in addition
the orbits are well aligned with the direction to the center
of the galaxy a feature that arises naturally in MOND but 
not in Newtonian gravity. Intuitively the MOND alignment can be 
understood as follows. In secular perturbation theory we think of
the orbits as wires with a non uniform density. The perihelion
is the lighter end of the wire and the aphelion the heavy
end. We see that the orbits are aligned with the heavy end
towards the galactic center consistent with the idea that
the orbits are responding to the galactic field. 

We now quantify the degree of alignment but before that 
it is worth noting here that in our dynamical analysis we
used a different coordinate system. The reference plane was perpendicular
to $\hat{{\bf n}}_G$ the unit vector that points in the direction of the 
center of the galaxy. We took the y axis to point in the
direction of $\hat{{\mathbf n}}_G$ and the reference plane to be the $z$-$x$ plane.
In the reference configuration the orbit lay in the $z$-$x$ plane
with the perihelion along the positive $z$ axis. The actual 
configuration is obtained by a $R_y$-$R_z$-$R_y$ sequence of rotations
by $\omega_K$, $i_K$ and $\Omega_K$. We distinguish the two systems notationally
by using upper case $(X,Y,Z)$ for the ecliptic frame and lowercase
$(x,y,z)$ for the frame that we used in the dynamical analysis. 
For the orientation variables we write $(\omega, i, \Omega)$ without subscripts
in the ecliptic frame and with subscripts $(\omega_K, i_K, \Omega_K)$ in 
the frame used for the dynamical analysis. 

In order to quantify the alignment it is useful first to calculate
the apsidal vectors $\hat{{\boldsymbol \alpha}}_K$ in the $XYZ$ frame. We note that 
$\hat{{\boldsymbol \alpha}}_K$
points along the positive $X$ axis when the orbit is in the reference
orientation. After being rotated to its actual orientation 
specified by $(\omega, i, \Omega)$ the vector $\hat{{\boldsymbol \alpha}}_K$ is given by
\begin{equation}
\hat{{\boldsymbol \alpha}}_K = 
\left( 
\begin{array}{c}
\cos \omega \cos \Omega - \sin \omega \sin \Omega \cos i \\
\cos \omega \sin \Omega + \sin \omega \cos \Omega \cos i \\
\sin \omega \sin i
\end{array}
\right)
\label{eq:alphadata}
\end{equation}
One way to quantify the alignment of the apsidal vectors with
$\hat{{\mathbf n}}_G$ is to compute the mean vaue of 
$\hat{{\boldsymbol \alpha}}_K \cdot \hat{{\mathbf n}}_G$ for the six
KBOs of interest. We obtain the value $-0.68$ for this alignment parameter.
Note that the alignment parameter is negative reflecting that the
apsidal vectors are anti-aligned with $\hat{{\mathbf n}}_G$. 
In the absence of MOND there should be no correlation 
between $\hat{{\mathbf n}}_G$ and the apsidal vectors and the expected
value of the alignment parameter is zero. If we take as
the null hypothesis that the apsidal vectors should be
independent unit vectors uniformly distributed over the
unit sphere then the observed value of the alignment parameter
is three standard deviations away from the expected value of zero. 
Another way to quantify the alignment of the apsidal vectors
is to note that the six values of $\hat{{\boldsymbol \alpha}}_K \cdot \hat{{\mathbf n}}_G$ 
lie in the range $-1$ to $-0.3$. The probability of that happening by chance is approximately
1 in 500. Thus we see that the degree of alignment is 
highly unlikely to be due to chance. 
Finally in Table 2 we enumerate the values of $A_K$, the angle
between $\hat{{\boldsymbol \alpha}}_K$ and $\hat{{\mathbf n}}_G$, 
for all six KBOs. In the preceding section we had estimated that 
these angles should lie between $129^\circ$ and an upper bound 
that depends on $a_K$. That upper bound is also listed in Table 2
for each KBO. Considering the crudeness of the estimated bounds the observed
values are in reasonable agreement with the predicted range. 

We can quantify the alignment of the apsidal vectors with $\hat{{\mathbf n}}_G$
by asking what is the probability that the alignment parameter 
(the mean value of $\hat{{\boldsymbol \alpha}}_{K} \cdot \hat{{\mathbf n}}_G$)
would be less than the observed value of $-0.68$ under the null hypothesis
that the apsidal unit vectors ${{\boldsymbol \alpha}}_K$ are independent and
uniformly distributed in orientation. A simple calculation shows that this
probability is 0.0011 or approximately one in a thousand. 

We can better quantify the alignment of the apsidal vectors with 
$\hat{{\mathbf n}}_G$ by using the Kolmogorov-Smirnov test as follows. 
We define
\begin{equation}
u_i = \frac{1}{2} ( 1 + \hat{{\boldsymbol \alpha}}_{Ki} \cdot \hat{{\bf n}}_G )
\label{eq:uidef}
\end{equation}
where $\hat{{\boldsymbol \alpha}}_{Ki}$ is the apsidal vector of the 
$i^{{\rm th}}$ Kuiper belt object (KBO) in the dataset. $ i = 1, 2, \ldots, n$ where
$n=6$ if we only consider the six KBOs that are known to have stable orbits.
Our prediction is that the $u_i$ values should cluster close to zero (since zero corresponds
to perfect anti-alignment between $\hat{{\boldsymbol \alpha}}_{Ki}$ and 
$\hat{{\bf n}}_G$). By contrast the null hypothesis is that the $u_i$ values 
are uniformly distributed random variables over the range $0 \leq u_i \leq 1$
(this follows from the hypothesis that the $\hat{{\boldsymbol \alpha}}_{Ki}$
are uniformly distributed over the unit sphere.)

\begin{figure}[h]
\begin{center}
\includegraphics[width=0.45\textwidth]{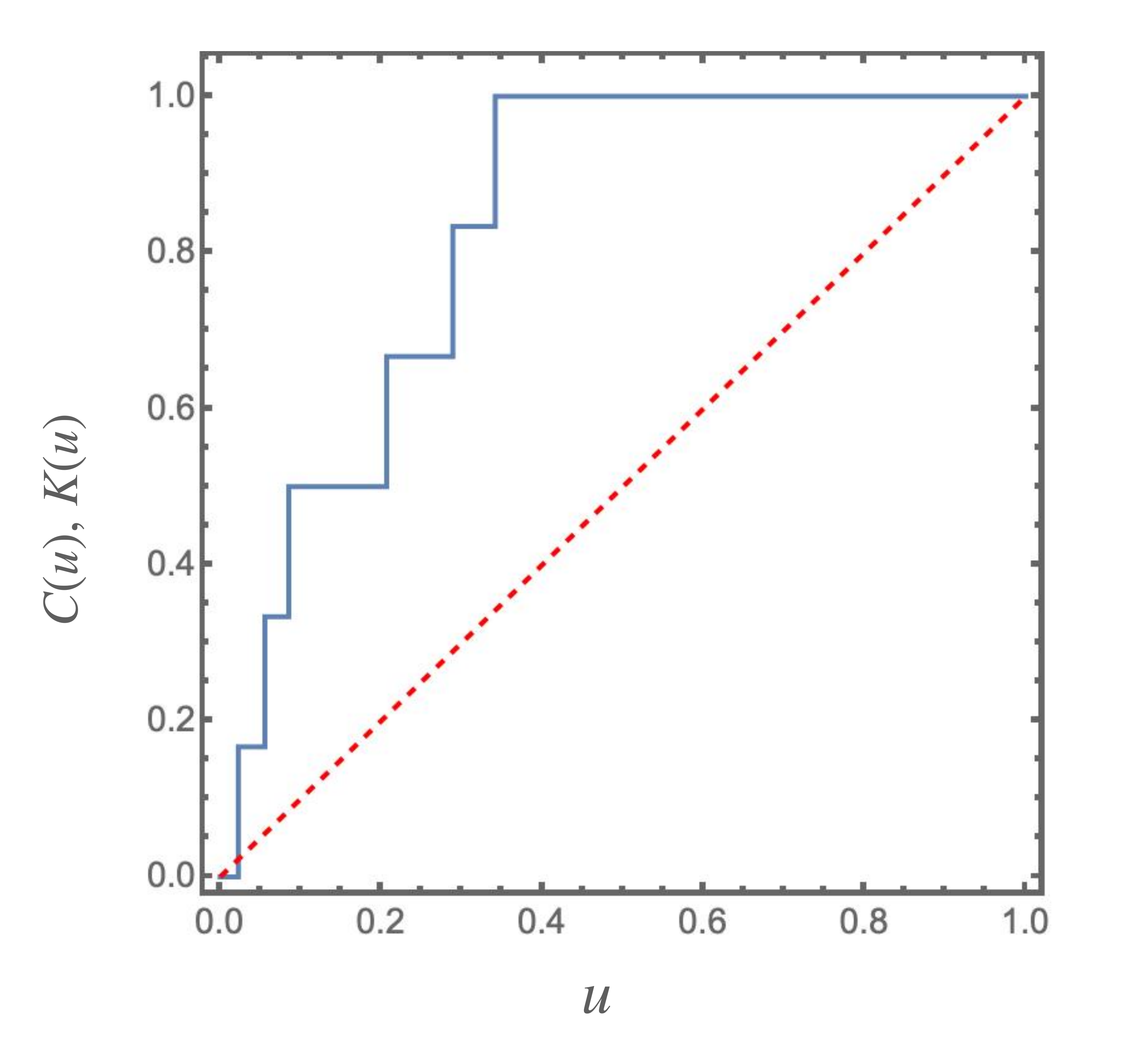}
\end{center}
\caption{Plot of the $C(u)$ staircase (blue) for the six KBOs of the Sedna family
that are known to have stable orbits. Also shown is $K(u)$, the cumulative
distribution function for the null hypothesis (red dashed line).}
\label{fig:ks6kbos}
\end{figure}

To qualitatively compare the data to the null hypothesis we plot the 
staircase
\begin{equation}
C(u) = \sum_{i=1}^n \Theta ( u - u_i )
\label{eq:staircase}
\end{equation}
where $\Theta$ denotes the step function [$\Theta (x) = 1$ for 
$ x >0 $ and $\Theta (x) = 0$ otherwise]. $C(u)$ should be compared to
\begin{equation}
K(u) = u
\label{eq:null}
\end{equation}
which is the cumulative distribution function corresponding to the null hypothesis.
Fig \ref{fig:ks6kbos} shows plots of $C(u)$ and $K(u)$ for the six stable KBOs;
we see that the data departs markedly from the null hypothesis. 

\vspace{2mm}
\noindent
To quantitatively compare the data to the null hypothesis we calculate the maximum vertical
deviation between the null cumulative distribution function and the data staircase,
\begin{equation}
\Delta = \max\limits_{0 \; \leq u \leq 1} | C(u) - K(u) |.
\label{eq:deltadef}
\end{equation}
We find the deviation for the case of the six stable objects $\Delta_6 = 0.658562\ldots$.
The Kolmogorov-Smirnov statistic is the probability that a deviation greater than
that observed will arise on the basis of the null hypothesis. Using the Kolmogorov-Smirnov
formula (see for example \cite{press} section 14-3)
we find that the probability of a deviation as large as $\Delta_6$ for $n=6$ is
$0.0054\ldots$ or approximately 1 in 200. 

\vspace{2mm}
\noindent
According to \cite{press} 
although the Kolmogorov-Smirnov 
formula is asymptotic for large $n$, it is quite accurate for $n$ as small as four. 
To confirm this we have calculated the Kolmogorov-Smirnov statistic by a second method.
We simulate a large number ($N = 300,000$) sets of data 
based on the null hypothesis. Each data set consists of $n$ random numbers drawn from
a uniform distribution over the unit interval. For each simulated data set we calculate
$\Delta$ and then we determine the fraction of data sets for which $\Delta$ exceeds the 
measured value $\Delta_n$. This fraction is an estimate of the Kolmogorov-Smirnov statistic. 
Using this method we obtain the values $0.0045\ldots$ for $n=6$.
We see that the Kolmogorov-Smirnov formula overshoots slightly,
but it is quite accurate even for $n=6$.
Hence the null hypothesis is falsified at a high level of significance. 

Thus we see that the apsidal vectors of the six KBOs are aligned with
the direction to the galactic center and this alignment is compatible 
with expectations based on MOND. To strengthen 
this conclusion we must look to future surveys to provide 
a larger observational sample and we need to refine the predictions of the MOND model
by carrying out large scale numerical simulations of KBO dynamics in MOND. 

In the literature the aspidal alignment of Sedna family
KBOs is sometimes quantified by reporting the clustering
of their longitudes of perihelion $\varpi = \omega + \Omega$ in the ecliptic
frame. As noted in the main text, 
we prefer to use $\hat{{\boldsymbol \alpha}}_K \cdot \hat{{\bf n}}_G$ 
as a measure of alignment
because it is frame invariant and unambiguously a measure
of alignment. However we have verified that the clustering
predicted by our model in $\omega_K$ and $i_K$ (together with complete
randomization in $\Omega_K$) does translate into a clustering in 
$\varpi$ for orbits with $i < 40^\circ$. 

Table 2 also includes the calculated value of the scaled
axial angular momentum $h$ for each KBO along with the lower
bound $h_{{\rm min}}$. We explain below how $\cos i_K$ can 
be computed from the known orbital data using eq (\ref{eq:cosikdata}). 
Using the computed values of $\cos i_K$, the tabulated values of $e_K$
and eq (\ref{eq:hdefined}) we can compute the values of 
$h$ for each KBO. $h_{{\rm min}}$ can be computed using
eq (\ref{eq:hmin}) and the tabulated values of $a_K$. We
see that the $h$ data lie in the range $h_{{\rm min}} < | h | < 
\sqrt{3/5}$ and generally towards the lower end of it as expected
for stability in the MOND analysis. 

\begin{figure}
	\begin{center}
		\includegraphics[width=4.0in]{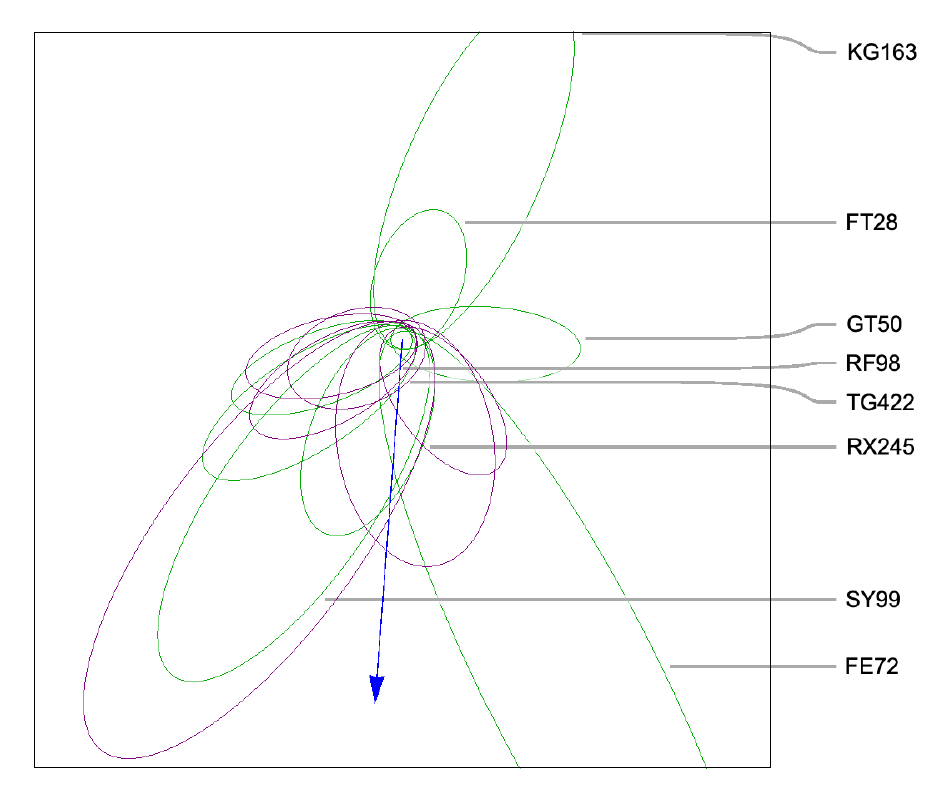}
		\caption{(Color online) Orbits of 14 KBOs of the Sedna family projected onto the ecliptic plane. The blue line is parallel to the projection of $\hat{{\bf n}}_G$ onto the ecliptic plane; it points towards the center of the galaxy. The orbits of the six KBOs that are stable are shown in purple; these are the same orbits plotted in fig 3 of the paper. Eight additional KBOs with metastable orbits are shown in green. These fourteen KBOs are the ones considered in support of the planet nine hypothesis in the review by \cite{review}. The orbital elements are from the Minor Planet Database and are given in Table I. 
}
		\label{fig:14align}
	\end{center}
\end{figure}

\begin{figure}
	\begin{center}
		\includegraphics[width=4.0in]{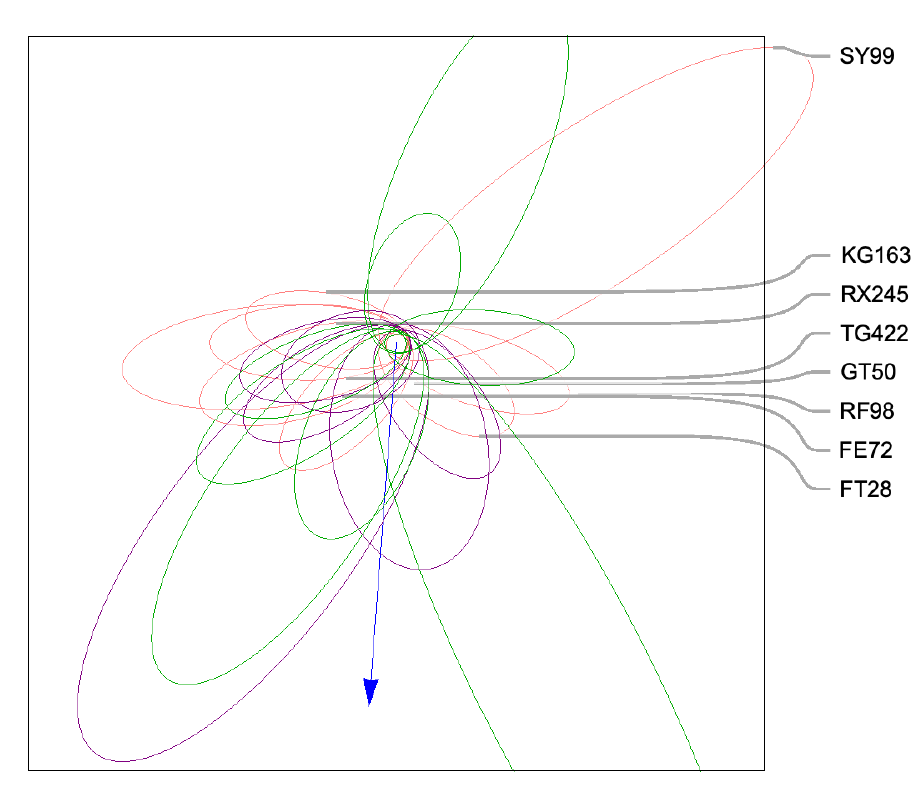}
		\caption{(Color online) Same as fig \ref{fig:14align} with eight additional KBO orbits plotted (pink curves). These are the KBOs that have been added to the Minor Planet Database since the publication of the review by \cite{review}. The fourteen objects plotted in fig \ref{fig:14align} are shown in purple (stable orbits) and green (metastable orbits). 
}
		\label{fig:22align}
	\end{center}
\end{figure}

\begin{figure}[h]
\begin{center}
\includegraphics[width=0.6\textwidth]{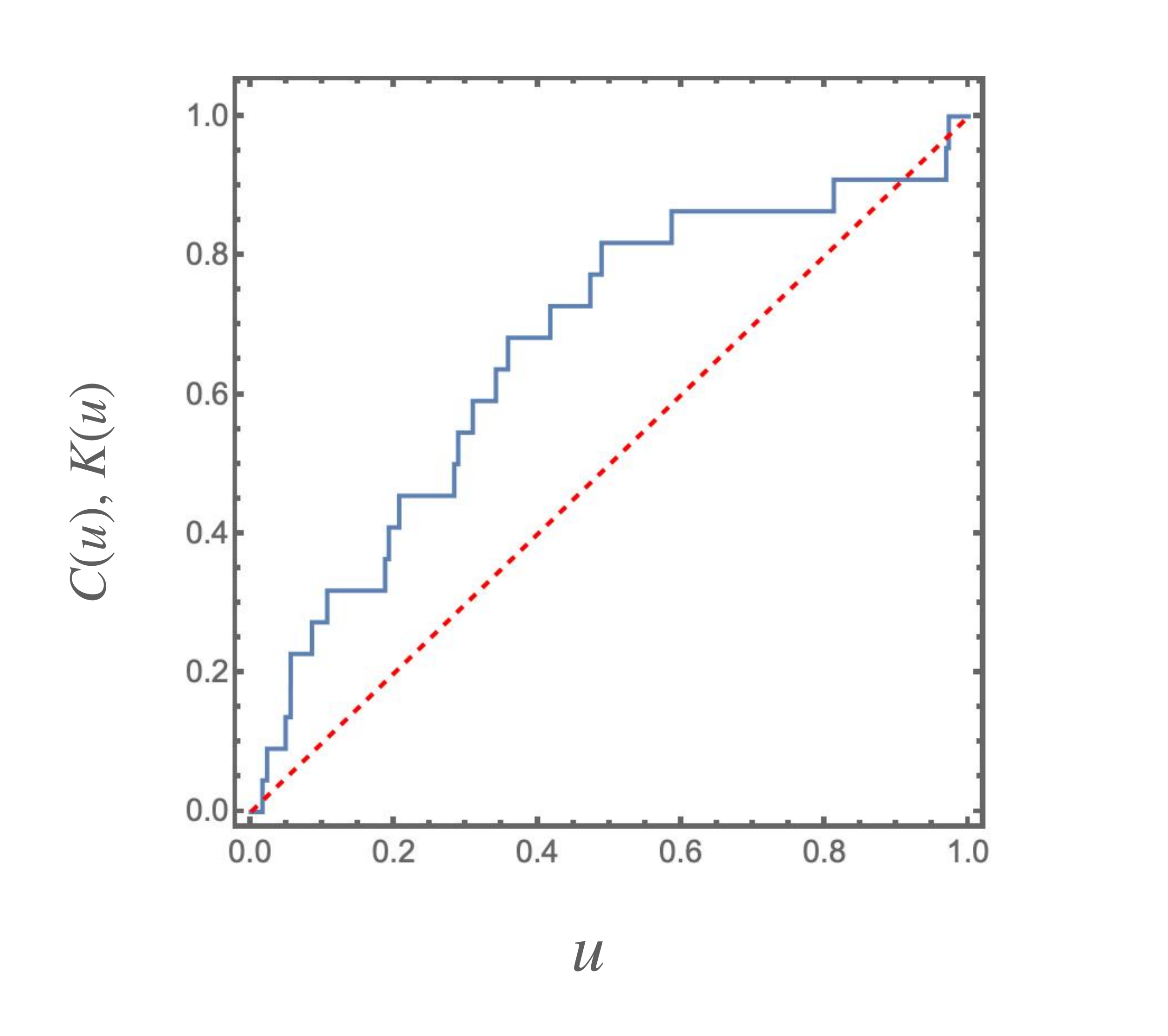}
\end{center}
\caption{Plot of the $C(u)$ staircase (blue) for all twenty two KBOs listed in 
Table 1. These objects meet the
criteria used in the review article by \cite{review} to identify potential members of the Sedna
family ($a \geq 250$ au, $q \geq 30$ au and $i < 40^\circ$). 
Also shown is $K(u)$, the cumulative
distribution function for the null hypothesis (red dashed line).}
\label{fig:ks22kbos}
\end{figure}

Thus far we have focused on the six Sedna family members with stable orbits, the best exemplars of this class. Fig \ref{fig:14align} shows the orbits of all 14 KBOs discussed in the review by \cite{review}. Fig \ref{fig:22align} shows the orbits of all 22 KBOs in Table I including the eight new ones that have been tabulated in the minor planet database since the review of \cite{review}. As in figure 3 of the paper the orbits and the direction to the galactic center are projected into the ecliptic plane. We see that the alignment between the orbits and the direction to the galactic center persists in this larger dataset. Eighteen of the twenty two orbits are well aligned. 
We can quantify this more precisely using the Kolmogorov Smirnov test. 
Fig \ref{fig:ks22kbos} shows the staircase corresponding to all twenty two objects
enumerated in Table I. The deviation of the staircase from a uniform distribution expected
on the basis of the null hypothesis is quite noticeable and we find
$\Delta_{22} = 0.329609\ldots$. The corresponding Kolmogorov Smirnoff
probability is $0.0125\ldots$ using the formula and $0.0123\ldots$
based on our Monte Carlo evaluation described above. 
Thus the null hypothesis that there is no alignment between the apsidal vectors and
$\hat{{\mathbf n}}_G$
can be ruled out at a high level of significance. 
As before the alignment of the orbits is consistent with both planet nine and MOND hypotheses, but the additional alignment with the galactic center provides further support for MOND.

\begin{table}
 \begin{ruledtabular}
 \begin{tabular}{lcccc}
 Object & $A_K$ & $A_{{\rm max}}$ & $h$ & $h_{{\rm min}}$ \\
 Sedna & 166$^\circ$ & 177$^\circ$ & $-0.14$ & 0.003\\
 TG387$^a$ & 151$^\circ$ & 179$^\circ$ & $0.0069$ & 0.0007\\
 2012 VP$_{113}$ & 117$^\circ$ & 175$^\circ$ & $-0.085$ & 0.01 \\
VN112$^b$ & 128$^\circ$ & 176$^\circ$ & 0.033 & 0.007\\
GB174 & 147$^\circ$ & 176$^\circ$ & $-0.18$ & 0.006 \\
SR349 & 110$^\circ$ & 176$^\circ$ & $0.081$ & 0.008 \\
 \end{tabular}
 \end{ruledtabular}
 \caption{\label{kbodatatoo}  Alignment angle $A_K$ and scaled axial angular momentum
 $h$ for six KBOs of the Sedna family compared to estimated MOND upper bound on
 $A_K$ and lower bound on $|h|$. 
 $^a$ TG387 is named
 Leleakuhonua. $^b$ VN112 is named Alicanto. }
 \end{table}

Finally we can use the data in Table 1 to compute the values
of $\omega_K$ and demonstrate that the six KBOs are clustered
in the $(e_K, \omega_K)$ phase space and moreover the clusering
is close to the approximate location of the fixed point. 
To this end it is useful to first compute the orbit
normal $\hat{{\mathbf l}}_K$ in the $XYZ$ frame. 
When the orbit is in the reference orientation $\hat{{\mathbf l}}_K$
is along the positive $Z$ axis. After the orbit it brought to 
the appropriate orientation $\hat{{\mathbf l}}_K$ is given by
\begin{equation}
\hat{{\mathbf l}}_K = 
\left(
\begin{array}{c}
\sin \Omega \sin i \\
- \cos \Omega \sin i \\
\cos i
\end{array}
\right)
\label{eq:lhatdata}
\end{equation}
We see from eq (\ref{eq:lhatdata}) that 
$\hat{{\mathbf l}}_K \cdot \hat{{\mathbf Z}} = \cos i$.
The corresponding formula for the frame used in the dynamical analysis is
\begin{equation}
\cos i_K = \hat{{\mathbf l}}_K \cdot \hat{{\mathbf n}}_G.
\label{eq:cosikdata}
\end{equation}
Since $0 \leq i_K \leq \pi$ it follows that $\sin i_K \geq 0$
and $\sin i_K$ can be unambiguously calculated from $\cos i_K$. 
Next we observe that according to eq (\ref{eq:alphadata}) 
$\hat{{\boldsymbol \alpha}}_K \cdot \hat{{\mathbf Z}} = \sin \omega \sin i$.
It follows that 
\begin{equation}
\sin \omega_K \sin i_K = \hat{{\boldsymbol \alpha}}_K \cdot \hat{{\mathbf n}}_G.
\label{eq:sinwkdata}
\end{equation}
By a similar argument we can also show
\begin{equation}
\cos \omega_K \sin i_K = ( \hat{{\mathbf l}}_K \times \hat{{\boldsymbol \alpha}}_K)
\cdot \hat{{\mathbf n}}_G 
\label{eq:coswkdata}
\end{equation}
Note that the right hand sides of eqs (\ref{eq:sinwkdata}) and
(\ref{eq:coswkdata}) can be computed from the known forms of 
all three vectors in the ecliptic frame. Together eqs (\ref{eq:sinwkdata})
and (\ref{eq:coswkdata}) unambiguously determine $\omega_K$.
It may be worth remarking both $i_K$ and $\omega_K$ are
fixed unambiguously once we have chosen the reference plane
to be perpendicular to $\hat{{\mathbf n}}_G$, the direction 
to the center of the galaxy. This choice is dictated by the physics
and is not arbitrary. However our choice of axes in the reference
plane is arbitrary and $\Omega_K$ will be dependent on that choice. 
Hence it is not meaningful to speak of clustering in $\Omega_K$
or $\varpi_K = \omega_K + \Omega_K$,
but it is meaningful to examine clustering in $\omega_K$, as we
now proceed to do. 

\begin{figure}
	\begin{center}
		\includegraphics[width=4.0in]{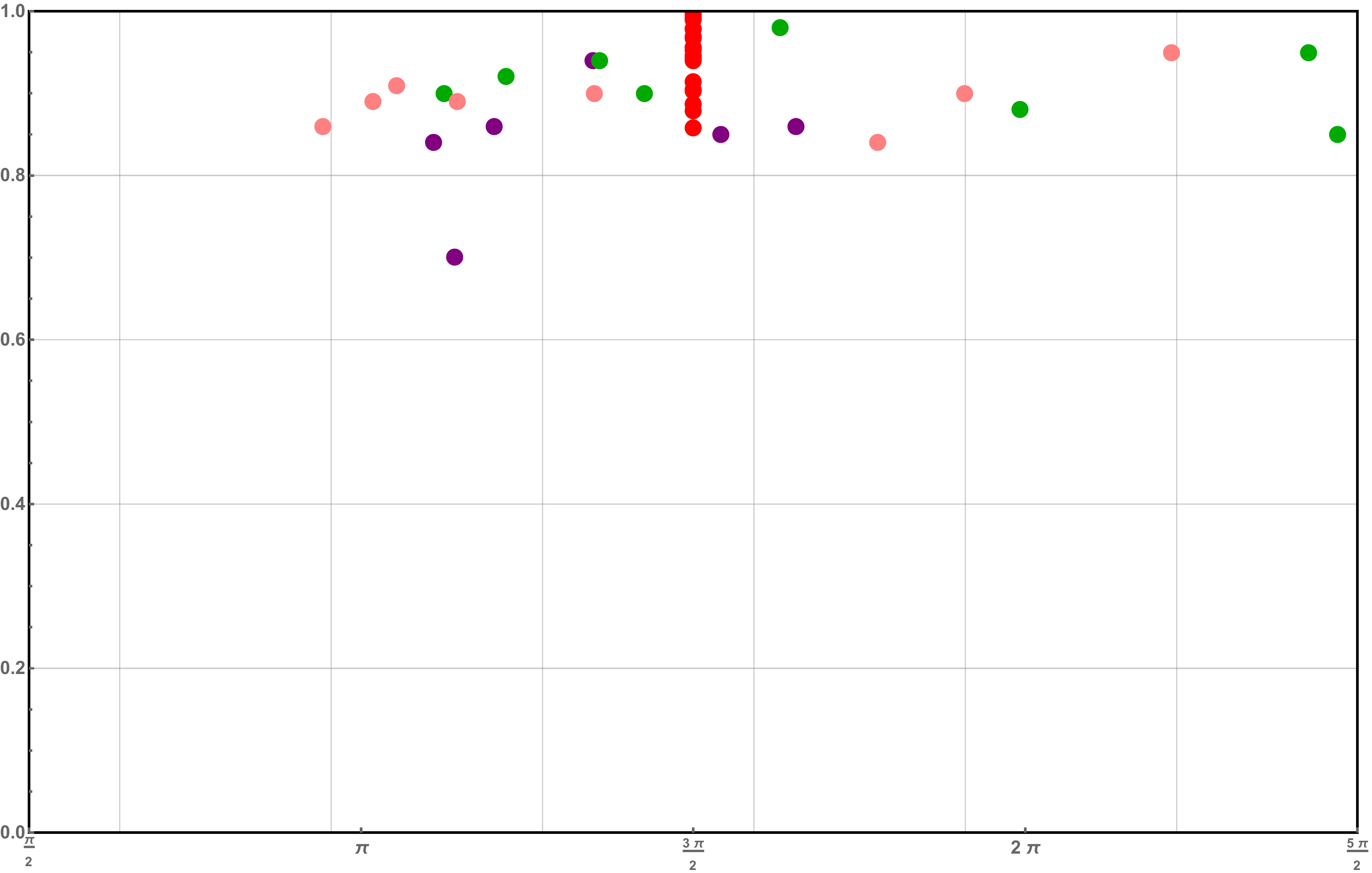}
		\caption{Plot of $( e_K^2, \omega_K^2 )$ for all 22 KBOs of the Sedna family listed in Table I (purple points for the six KBOs known to have stable orbits, green points for the ones with metastable orbits, and pink points for the eight added to the minor planet database since the publication of the review by \cite{review}). Also shown is the location of the corresponding fixed point of the MOND orbital dynamics in the quadrupole approximation (red points). The clustering shown in Fig 4 of the main body of the paper for the six KBOs with known stable orbits is seen to persist for all twenty two KBOs.}
		\label{fig:phase22}
	\end{center}
\end{figure}

Fig. 4 in the main body of the paper
shows the clustering of the six KBOs in 
the $(e_K, \omega_K)$ plane. The $\omega_K$ values were calculated
using eqs (\ref{eq:cosikdata}), (\ref{eq:sinwkdata}) and (\ref{eq:coswkdata})
and the $e_K$ values are the ones enumerated in Table 1. These points
are shown in blue. Also shown in red in the same plot are the locations
of the corresponding fixed points $(e_C, 3 \pi/2)$ in the quadrupole
approximation. We see that the Sedna family of KBOs is clustered 
close to the fixed points in phase space as predicted. 
Fig \ref{fig:phase22} in this section shows that the clustering in phase space persists when we include all twenty two KBOs enumerated in table I: the six with stable orbits shown in the main body of the paper as well as the eight with metastable orbits and the eight that have appeared in the Minor Planet Database since the publication of the review by \cite{review}.

To strengthen the evidence for phase space clustering we must look
to future surveys to provide a larger observational sample of KBOs
of the Sedna family. Also numerical simulations of the dynamics will
allow more precise location of the fixed point and its domain of
stability. These calculations are underway.

\end{document}